\documentclass[fleqn,usenatbib]{mnras}

\usepackage{newtxtext,newtxmath}

\usepackage[T1]{fontenc}
\usepackage{ae,aecompl}

\usepackage{graphicx}	
\usepackage{amsmath}	
\usepackage{amssymb}	

\newcommand{\livrevrel}{Living Rev. Relativ.}
\newcommand{\revmodphys}{Rev. Mod. Phys.}
\newcommand{\advspres}{Adv. Space Res.}
\newcommand{\intjmodphysd}{Int. J. Modern Phys. D}
\newcommand{\intjmodphysa}{Int. J. Modern Phys. A}
\newcommand{\cosmres}{Cosmic Res.}
\newcommand{\pnas}{Proc. Natl. Acad. Sci.}
\newcommand{\genrelgrav}{Gen. Relativ. Gravitation}
\newcommand{\advastron}{Adv. Astron.}
\newcommand{\ab}{Astrophys. Bull.}
\newcommand{\astrep}{Astron. Rep.}
\newcommand{\ARAAAJ}{ARA\&A}
\newcommand{\pasaust}{Publ. Astron. Soc. Aust.}

\title[High-Redshift LGRB Hubble Diagram]
{ High-Redshift Long Gamma-Ray Bursts  Hubble Diagram as a Test of Basic Cosmological Relations}

\author[S. I. Shirokov et al.]{
S. I. Shirokov,$^\text{1}$\thanks{E-mail: arhath.sis@yandex.ru}
I. V. Sokolov,$^\text{2}$
N. Yu. Lovyagin,$^\text{3}$
L. Amati,$^\text{4}$
Yu. V. Baryshev,$^\text{3}$
\newauthor{V. V. Sokolov,$^\text{5}$
and V. L. Gorokhov$^\text{6}$}
\\
$^\text{1}$SPb Branch of Special Astrophysical Observatory of Russian Academy of Sciences, 65 Pulkovskoye Shosse, St Petersburg 196140, Russia\\
$^\text{2}$Institute of Astronomy of the Russian Academy of Sciences, Pyatnitskaya Str 4, Moscow 119017, Russia\\
$^\text{3}$Saint Petersburg State University,
7/9 Universitetskaya Nab., St Petersburg 199034, Russia\\
$^\text{4}$INAF -- Istituto di Astrofisica Spaziale e Fisica Cosmica, di Bologna, Via Gobetti 101, I-40129 Bologna, Italy\\
$^\text{5}$Special Astrophysical Observatory,
Nizhnij Arkhyz, Zelenchukskiy Region, Karachai-Cherkessian Republic 369167, Russia\\
$^\text{6}$Saint Petersburg Electrotechnical Univeristy,
Ulitsa Professora Popova 5, 197376 St. Petersburg, Russia
}

\date{Accepted XXX. Received YYY; in original form ZZZ}

\pubyear{2020}

\begin{document}

\label{firstpage}
\pagerange{\pageref{firstpage}--\pageref{lastpage}}
\maketitle

\begin{abstract}
We examine the prospects of the high-redshift long gamma-ray bursts (LGRBs) Hubble diagram  as a test of the basic cosmological principles. Analysis of the Hubble diagram  allows us to test several fundamental cosmological principles using the directly observed flux--distance--redshift relation.
Modern LGRBs data together with the correlation between the spectral peak energy and the isotropic equivalent radiated energy (the so-called Amati relation)
can be used for construction of the Hubble diagram at the model-independent level. 
We emphasize observational selection effects, which inevitably exist and  distort the theoretically predicted relations. 
An example is the weak and strong gravitational lensing bias effect for high-redshift LGRB in the presence of limited observational sensitivity (Malmquist bias).
After bias correction, there is a tendency to vacuum-dominated models with
$\Omega_\Lambda \rightarrow 0.9$, 
$\Omega_\text{m} \rightarrow 0.1$.
Forthcoming gamma-ray observations by the Transient High-Energy Sky and Early
Universe Surveyor (THESEUS) space mission together with ground and space-based multimessenger facilities will allow us to improve essentially the restrictions on alternative basic principles of cosmological models.
\end{abstract}
%



\section{Introduction}

The Hubble diagram (HD)  is the directly observed flux--distance--redshift relation for a sample of ``standard candles''. It was the first classical observational cosmological test performed in 1929 by Hubble in his classical paper~\citet{Hubble1929}. The observable redshift interval was 0--0.003, i.e. galaxy distances $d  < 15$ Mpc.  This test led to the discovery of the fundamental cosmological  observational  ``Hubble law'' --  the linear  redshift--distance relation $z = Hd/c$ by using the Euclidean flux--distance relation $F=L/4\pi d^2$ for the standard candle. The linearity of this Hubble law on the scales
1--300 Mpc was confirmed by Sandage at the Hale Telescope~\citep*[current results are presented in][]{Sandage2010}.

70 yr later, the HD was constructed for supernova Type Ia (SN Ia) standard candles up to 
$z\approx1$ and, surprisingly, the accelerated expansion of the Universe within the standard Friedmann--Lemaitre--Robertson--Walker (FLRW) cosmological model was discovered. This test led in 1998 to the introduction of the dark energy into the standard cosmological model (SCM) by Riess's and Perlmutter's teams~\citep{Riess1998,Perlmutter1999}.

Nowadays the observational cosmology
is based on the multimessenger astronomy that combines observations of electromagnetic radiation (from radio to gamma-ray bands), cosmic rays, neutrino and gravitational waves. In particular, the Transient High-Energy Sky and Early Universe Surveyor (THESEUS) space mission project is aimed to  explore the unique capabilities of gamma-ray bursts (GRBs) for cosmology and multimessenger astrophysics
\citep{Amati2018, Strata2018}. 
Because of the huge GRB luminosity, the THESEUS  will bring crucial data for testing different theoretical predictions of cosmological models up to redshifts 
$z \sim 10$. 

Here we use  data for the long gamma-ray bursts (LGRB) with known redshifts and study the prospects of applying the HD for testing  basic cosmological predictions, such  as redshift--distance and flux--distance theoretical relations for high redshifts. 
Considering LGRB
as a standard candle, one uses observed luminosity correlations, such as the Amati relation between the
energy (frequency) of the spectral peak $E_\text{p}$ and isotropic equivalent radiated energy $E_\text{iso}$
\citep{Amati2002,Amati2008,Amati2019,Demianski2017a,Demianski2017b,Lusso2019}. 
Though there are correlations of GRB luminosity with other observed GRB parameters \citep[e.g.,][]{Yonetoku2004,  Wang2011a, Wei2017},
here we consider the Amati relation as the simplest (less model dependent) among all GRB correlations investigated for cosmology.

One difficulty in the analysis of the HD using high-redshift LGRBs is the cosmologically model-independent calibration 
of the GRB luminosity~\citep{Kodama2008, Liang2008, Demianski2017a, Demianski2017b, Amati2019}. 
In order to build a model-independent HD at redshifts $z < 1$ we use the recent sample of SNe Ia~\citep{Scolnic2018}.

Very important additional obstacles come from  different observational biases, selection and evolution effects, especially the Malmquist bias (MB) and the gravitational lensing bias (GLB),  which distort the observed LGRB parameters and, hence, the true LGRB luminosity
~\citep{Wang2011b,Deng2016, Kinugawa2019,Lloyd2019,Xue2019}.

Besides the weak gravitational lensing
~\citep[GL;][]{Wang2011b} 
we consider the effective bias due to the Malmquist selection effect and  the strong GL within the fractal matter distribution. According to modern observational data the strong GL plays an important role in cosmology
\citep{Acebron2020,CervantesCota2020,Shajib2019}
and we study its possible influence 
on the LGRB HD.

In Sec.~\ref{Hubble Diagram in Cosmological Models}, we derive the basic cosmological relations between theoretical and observable quantities that are used in the HD construction. 
In Sec.~\ref{LGRB standard candles and Amati relation}, we consider how the LGRBs are used as standard candles, and in 
Sec.~\ref{SN Model Independent Hubble Diagram}, we construct the model-independent HD for the SNe Ia sample at redshifts 
$z<1.4$.
 Sec.~\ref{LGRB Hubble Diagram for different cosmological models}
is devoted to an analysis of the high-redshift LGRB HD for different cosmological models, taking into account the GLB.
For illustration of our approach we consider the LGRBs sample~\citep{Amati2019} and compare the results of our HD fittings.
We discuss the results and give our conclusions in
Sec.~\ref{Discussion and conclusions}.

\section{Hubble Diagram in Cosmological Models}
\label{Hubble Diagram in Cosmological Models}

Modern physics considers the observable Universe as a part of ``the cosmic laboratory'', where all basic principles and main physical fundamental laws can be tested with increasing accuracy. In particular, in the spirit of the modern theoretical physics, such cosmological basis as the constancy of fundamental constants, the equivalence principle, the Lorentz invariance, the cosmological principle, the general relativity and its modifications, and the space expansion paradigm must be tested in the cosmic laboratory ~\citep[e.g.,][]{Turner2002, Uzan2003,  Baryshev2012, Baryshev2015, Clifton2012, deRham2017, Amendola2018, Ishak2018, Lusso2019, Perivol2019, Riess2019}.

The HD is a necessary (but not sufficient) cosmological test that allows one to select some basic initial theoretical postulates of cosmological models.
Following the practical cosmology approach~\citep[see][]{Sandage1995,Baryshev2012},
we  construct the HD through the Amati relation based on recent GRB catalogues. We compare the observed HD with theoretical predictions of several cosmological models. For testing particular cosmological assumptions and for demonstration of the ability of the HD test, we consider several examples of the cosmological redshift--distance--flux relations: general Friedmann's space expansion models, including $\Lambda$ cold dark matter ($\Lambda$CDM) and quintessence  $w$CDM, classical steady-state, and also the field-fractal model in Minkowski space, and the tired-light model in Euclidean space.

\subsection{Observed quantities for the Hubble diagram}
In cosmology it is very important to distinct between experimentally measured and theoretically inferred cosmological relations (e.g. the observed flux--redshift law and the theoretical expansion velocity--redshift law). In the HD the directly measured quantities are flux and redshift, which  must be considered as primary physical quantities for all cosmological models.
Hence this practical cosmology test considers the correlation between two quantities measured on the Earth -- the observed  frequency (wavelength) shift $z$ and the observed spectral energy flux 
$F_\nu$, calculated for an object with known spectral luminosity $L_\nu$ (``standard candle'').

The  direct empirical cosmological quantity is the redshift $z$, which is defined as 
\begin{equation}
\label{redshift}
 z = \frac{\lambda_\text{obs} -\lambda_\text{lab}}{\lambda_\text{lab}}
 = \frac{\lambda_\text{obs}}{\lambda_\text{lab}} - 1 =
 \frac{\nu_\text{source}}{\nu_\text{obs}} - 1
\,,
\end{equation}
where $\nu_\text{obs} = c/\lambda_\text{obs}$ is the  photon frequency observed 
at a telescope, and $\nu_\text{lab}$ is the  photon frequency measured in laboratory for the same physical process, which is assumed to be equal to the emitted photon frequency at the source $\nu_\text{source}=\nu_\text{lab}$ (universality of physical laws).

For small redshifts there is a spectroscopic tradition to express cosmological redshift in terms of the ``apparent radial velocity'' 
$V_\text{app}/c \ll 1$ by using definition
\begin{equation}
\label{app-vel-redshift}
 z = 
 \frac{\nu_\text{source}}{\nu_\text{obs}} - 1 
 \approx  \frac{V_\text{app}}{c}
\,.
\end{equation}

The second cosmological quantity directly observed at a telescope is the energy flux 
$F_L$ having the dimension (erg~$\cdot$~cm$^{-2}\cdot$~s$^{-1}$) measured within the interval $i = (\nu_1, \,\nu_2)$ of frequencies $\nu$ (energies $h\nu$). 
The flux $F_L$ is defined as the energy of photons crossing the unit area of a detector per second, which were emitted by a source with the luminosity $L(\nu)$ and reach a telescope at the ``luminosity distance'' $d_L$, that is
\begin{equation}
\label{flux}
  F_{L,i}
    = \frac{L_i}{4\pi d_L^\text{2}}\, ,
    \qquad 
 L_i = \int_{\nu_1}^{\nu_2} L(\nu)\text{d}\nu\,,
\end{equation}
where the comoving spectral luminosity (power) of the source $L(\nu)$ is measured in 
erg~$\cdot$~s$^{-1}\cdot$~Hz$^{-1}$. 
For $\nu_1 = 0$ and $\nu_2 = \infty$, we have bolometric luminosity $L$ measured in 
erg~$\cdot$~s$^{-1}$
and bolometric flux $F_L$. 

The integral energy received by a detector during the burst duration time is the ``fluence'' $F_E$ measured in units erg~$\cdot$~cm$^{-2}$ from a source at the ``energy distance'' $d_E$
\begin{equation}
\label{flux-E}
  F_E
    = \frac{E_\text{iso}}{4\pi d_E^\text{2}}\,,
    \qquad 
 E_\text{iso} = \int_{t_1}^{t_2} L_i(t)\text{d}t\,,
\end{equation}
where $E_\text{iso}$ (erg) is the total energy radiated isotropically by a given source during the burst time.

In order to use the HD as an observational cosmological test one needs to construct theoretical concepts, which correspond to the above measured quantities at the terrestrial (space) observatory. Here a cosmological model enters the discussion, and 
there are several theoretical redshift--distance--flux relations~\citep{Peacock1999, Baryshev2012, Clifton2012, Amendola2018, Ishak2018, Perivol2019}, which can be used for the HD construction. 

For each considered cosmological model a predicted theoretical HD contains two basic theoretical relations: distance--redshift and flux--distance.

\subsection{The standard FLRW model}
\label{standard-model-1}

Here we consider the theoretically inferred basic cosmological relations in the framework of the Friedmann--Lemaitre--Robertson--Walker model 
(or simply the Friedmann model).
In many textbooks on FLRW cosmological model there is no clear discussion of the physical and geometrical sense of the concept of ``space expansion'',  which leads to confusion between the Doppler and Lemaitre redshift effects 
\citep{Francis2007, Baryshev2015}.

So, it is important to present the definitions of the theoretical quantities needed for construction of the HD.

\subsubsection{Basic principles}
The standard Friedmann  model may be defined according to three  basic initial principles~\citep{Baryshev2012}. 

First, Einstein's 
general relativity theory (GRT) describes gravity by the metric tensor $g^{\mu \nu}$ 
of the curved Riemannian space, which obeys the field equations
\begin{eqnarray}
 \Re^{\mu \nu} - \frac{1}{2}g^{\mu \nu}\,\Re = \frac{8\pi G}{c^4}T^{\mu \nu}_\Sigma\,,
\label{einstein}
\end{eqnarray}
where $\Re^{\mu \nu}$ is the Ricci tensor, 
$\Re$ is the scalar curvature, and
$T^{\mu \nu}_{\Sigma}$ is the total energy--momentum tensor of all kinds of matter, including dark energy and dark matter (but it does not include the energy--momentum tensor of the gravity field itself).

Second, Einstein's cosmological principle (ECP)
states the strict mathematical homogeneity for the dynamically important matter, i.e. 
$\rho(\vec{r},t) = \rho (t)$, 
$p(\vec{r},t) = p(t)$, and
$g_{\mu \nu}(\vec{r},t) = g_{\mu \nu}(t)$. 
It is assumed that the observed inhomogeneity of visible distribution of matter does not influence the global homogeneous total matter distribution.

From ECP, the GRT field equation Eq.~(\ref{einstein}) give the Friedmann equation
\begin{equation}
\label{f1}
 \Omega -1 = \Omega_{k}\,,
  \qquad
 q = - \ddot{S}S/\dot{S}^2 =
 \frac{1}{2}\Omega \left(1 + \frac{3p}{\varrho c^2} \right)\,,
\end{equation}
where  $\Omega = \varrho / \varrho_\text{crit}$,
~$\varrho_\text{crit}=3H^2/8\pi G$,~ $\Omega_{k} = kc^2/S^2H^2$,  ~$ q = -\ddot{S}S/\dot{S}^2 $, and
~$H=\dot{S}/S$.
Here $\Omega, p ,\varrho$ are the total quantities for the sum of non-interacting substances, which obey the Bianchi identity.

\begin{figure}
    \centering
    \includegraphics[width=0.47\textwidth]{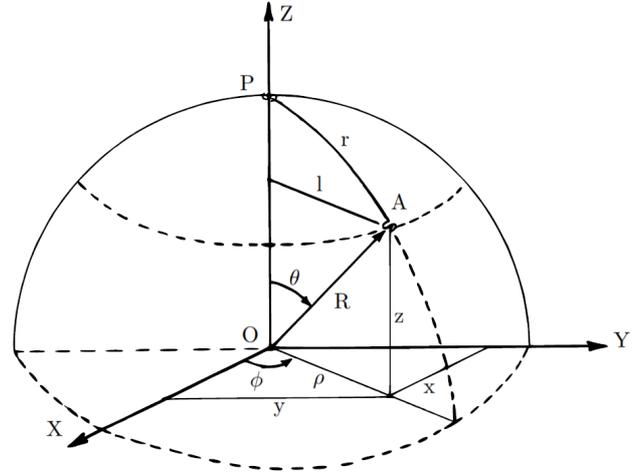}
    \caption{Geometrical sense of the line element  coordinates Eqs.~(\ref{rw1}) and~(\ref{rw2}) for the case of the 2D spherical expanding space embedded in the 3D Euclidean space. The radius $R$ of the 2D sphere is the scale factor $R(t) = S(t)$ of the expanding 2D space. The ``internal'' metric distance between the galaxies $P$ and $A$ is $r$ and the ``external'' metric distance is $l$. The angle $\theta$ is the dimensionless comoving ``distance''~$\chi$.
    }
    \label{fig:space}
\end{figure}

\begin{figure}
    \centering
    \includegraphics[width=0.47\textwidth]{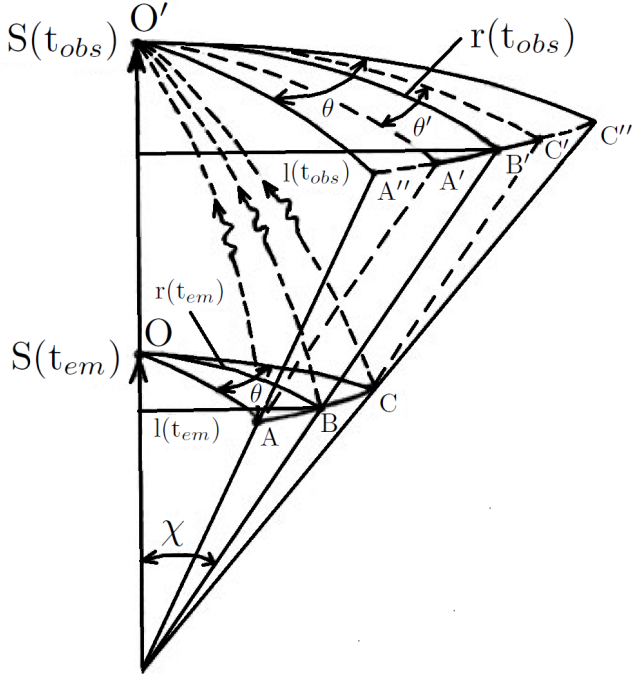}
    \caption{The 2D spherical expanding space embedded in the 3D Euclidean space. Metric distances at the times of emission $t_\text{em}$ and observation $t_\text{obs}$ of light from a distant galaxy have the linear size $ABC$. Note that $l(t_\text{em})$ and $r(t_\text{em})$ are the external and internal metric distances, respectively, when the scale factor is $S(t_\text{em})$.}
    \label{fig:space_props}
\end{figure}

Third, the Riemannian space is expanding and it is described by the time-dependent proper distance $r(t) = S(t)\chi$,   
where $S(t)$ is the scale factor and $\chi$ is the comoving distance. Thus the observed
cosmological redshift $z$ is given by
the Lemaitre effect of this space stretching
(this is not the Doppler effect),
\begin{equation}
 \label{lem}
 z = \frac{\nu_\text{source}}{\nu_\text{obs}} -1 =
 \frac{S(t_\text{obs})}{S(t_\text{em})} -1\,,   
\end{equation}
where $S(t_\text{em})$ and $S(t_\text{obs})$ are the scale factors at the times of emission and reception, respectively 
(Figs.~\ref{fig:space} and \ref{fig:space_props}). 

The fundamental consequence of the mathematical homogeneity and isotropy (ECP) is that the Einstein's GRT equations, Eq.~(\ref{einstein}),
has the Friedmann's form Eq.~(\ref{f1})
and the line elements of coordinates in d$s^2 = g_{\mu\nu}x^{\mu}x^{\nu}$
are given by  the standard FLRW form
in terms of the ``internal'' metric distance
$r(t) = S(t)\chi $
\begin{equation}
  ds^2 = c^2dt^2 - S^2(t)d\chi^2 - S^2(t) I_{k}^2 (\chi) (d\theta^2 +  \sin^2{\theta} d\phi^2)\,,
\label{rw1}   
\end{equation}
where 
$I_{k}(\chi) = (\sin(\chi),~\chi,~\sinh(\chi))$, for $k = (+1,~0,-1)$,  $r$ is the proper metric ``internal'' distance,  $S(t)$ is the scale factor with the dimension of length $[S]$ = cm, and $\chi$ is the dimensionless comoving ``distance''. 

In terms of the ``external'' metric distance 
$l(t) = S(t)\mu $ and dimensional comoving coordinate $\mu=I_{k}( \chi)$ 
(Fig.~\ref{fig:space}),
the interval d$s$ has the form
\begin{equation}
    \label{rw2}
  ds^\text{2} = c^\text{2}dt^\text{2} - S^\text{2}(t)
  \frac{d\mu^\text{2}}{1-k\mu^2} - S(t)^\text{2} \mu^\text{2}
 (d\theta^\text{2}+\sin^\text{2}\theta d\phi^\text{2})\,.
\end{equation}

The geometrical sense of these coordinates is given in Figs.~\ref{fig:space} and \ref{fig:space_props} for the case of the expanding 2D spherical space embedded in the 3D
Euclidean space. Here the radius $R$ of the 2D sphere is the scale factor $R(t) = S(t)$ and the angle $\theta$ is the dimensionless comoving ``distance''~$\chi$.

The relation between the internal and external distances
\begin{equation}
\label{r}
r(t) = S(t)\cdot\chi\,, 
\qquad
l(t) = S(t)\cdot \mu  
\end{equation}
can be written as
\begin{equation}
\label{r-l}
r = S(t) I^{-1}_\text{k}(l/S)\,, \qquad l=S(t)I_{k}(r/S)\,, 
\end{equation}
where $I^{-1}_{k}$ is the inverse function for $I_{k}$. 
In the case of the flat universe ($k=0$) these distances coincide $r(t)=l(t)$.

Note that the exact general relativity expression for the space expansion velocity $V_\text{exp}$ is the time derivative of the metric distance given by Eq.~(\ref{r}), so that
\begin{equation}
 \label{expvel}
 V_\text{exp}(r) = \frac{dr}{dt} =  \frac{dS}{dt}
 \chi = \frac{dS}{dt}\cdot \frac{r}{S} = H(t) r = c \frac{r}{R_{H}} \,.  
\end{equation}

The expression of $V_\text{exp}(z)$ is more complex and includes the distance--redshift relation $r(z)$. The derivation of the FLRW relations between metric$/$luminosity distances and redshift is given in Appendix B.

\subsubsection{Distance modulus, apparent and absolute  magnitudes}

The apparent stellar magnitude $m_i$ of an object  observed through a filter ``$i$'' is defined 
via the ratio of the measured flux $F_{L,i}$ to the standard flux $F_0$. Taking into account 
Eq.~(\ref{flux}) we can write
\begin{equation}
    m_i= -2.5\log(F_{L,i} /F_0)=
    5\log(d_L)+ 25 + M_i + C_i\,,
    \label{m-def-L}
\end{equation}
where  $d_L$ is the luminosity distance measured in Mpc,   $M_i$ is the absolute stellar magnitude of the standard source, and 
$C_i(z) =  K_i + A_i + E_i$ involves different observational corrections, where the $K_i,\,A_i$, and $E_i$  are redshift biases, extinction, and evolution corrections  known for a standard candle class.

For the case of the bolometric fluxes $F_L$ and fluences $F_E$ given by Eqs.~(\ref{flux}) and (\ref{flux-E}) we  define the corresponding observed ``distance modulus'' as the luminosity distance modulus
\begin{equation}
   \mu_L = \,
    m_{L} - M_{L} =
    5\log(d_L)+ 25 + C_{L,i}(z)\,,
    \label{mu-L}
\end{equation}
and the energy distance modulus
\begin{equation}
   \mu_{E} = \,
    m_{E} - M_{E} =
    5\log(d_E)+ 25 + C_{E,i}(z)\,.
    \label{mu-E}
\end{equation}

To construct the HD we will fit observed flux and fluence distance moduli by the theoretical ones, which are given by 
Eqs.~(\ref{lum-dist-L}) and (\ref{lum-dist-E}). Thus we get the HD as the ($\mu$ versus $z$)  relations 
in two forms: for the measured flux $F_L$,
\begin{equation}
   \mu_L = 
       5\log(l(z)(1+z))+ 25 =
       5\log l(z) + 5\log (1+z) +25\,,
    \label{mu-L-Frid}
\end{equation}
and for the measured fluence $F_E$,
\begin{equation}
   \mu_E = 
       5\log(l(z)(\sqrt{1+z})+ 25 =
       5\log(l(z) + 2.5\log (1+z) + 25\,.
    \label{mu-E-Frid}
\end{equation}

In fact for the SNe Ia observations we have measurements of the ``luminosity distance modulus'' $\mu_L(z)$. For GRB fluence, observations give the ``energy distance modulus'' $\mu_E(z)$. Hence, to construct the HD in terms of the GRB luminosity distance modulus  $\mu_L$, Eq.~(\ref{mu-L-Frid}), we must correct the GRB $\mu_E$ observational data by the factor
\begin{equation}
  \mu_L = \mu_E + 
       2.5\log (1+z)\,\,.
    \label{dmu-L-E-Frid}
\end{equation}
The correction Eq.~(\ref{dmu-L-E-Frid}) takes into account the additional factor $(1+z)$ in 
Eq.~(\ref{flux}), which relates to the time dilation effect  for the observed luminosity distance modulus of the GRBs.

\subsection{Testing basic cosmological assumptions}
\label{models-param}

Above, we considered three basic SCM principles, 
which include the geometrical gravity theory (general relativity), ECP
of the strict mathematical homogeneity for the dynamically important matter, and the space expansion paradigm for the observed cosmological redshift.

In the spirit of practical cosmology the basic initial principles of the SCM must be tested with increasing accuracy and compared with other cosmological models at wider redshift intervals
\citep{Baryshev2012,Baryshev2015}.

Here, for illustration of the HD predicted on the basis of different initial principles, besides SCM, we also
consider the CSS, FF and TL models.

\begin{figure*}
    \includegraphics[height=0.49\textwidth,angle=-90]{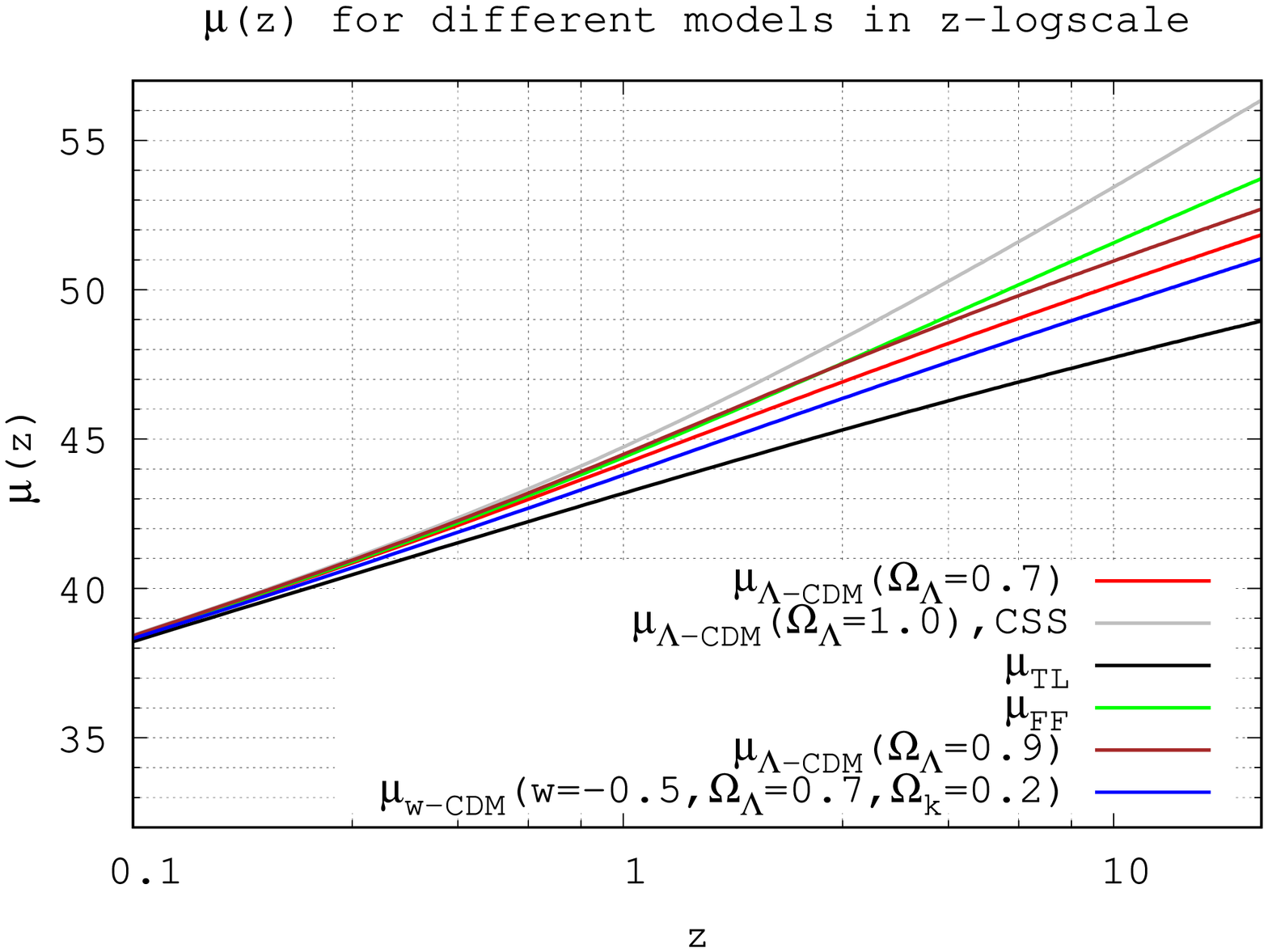}
    \hfill
    \includegraphics[height=0.49\textwidth,angle=-90]{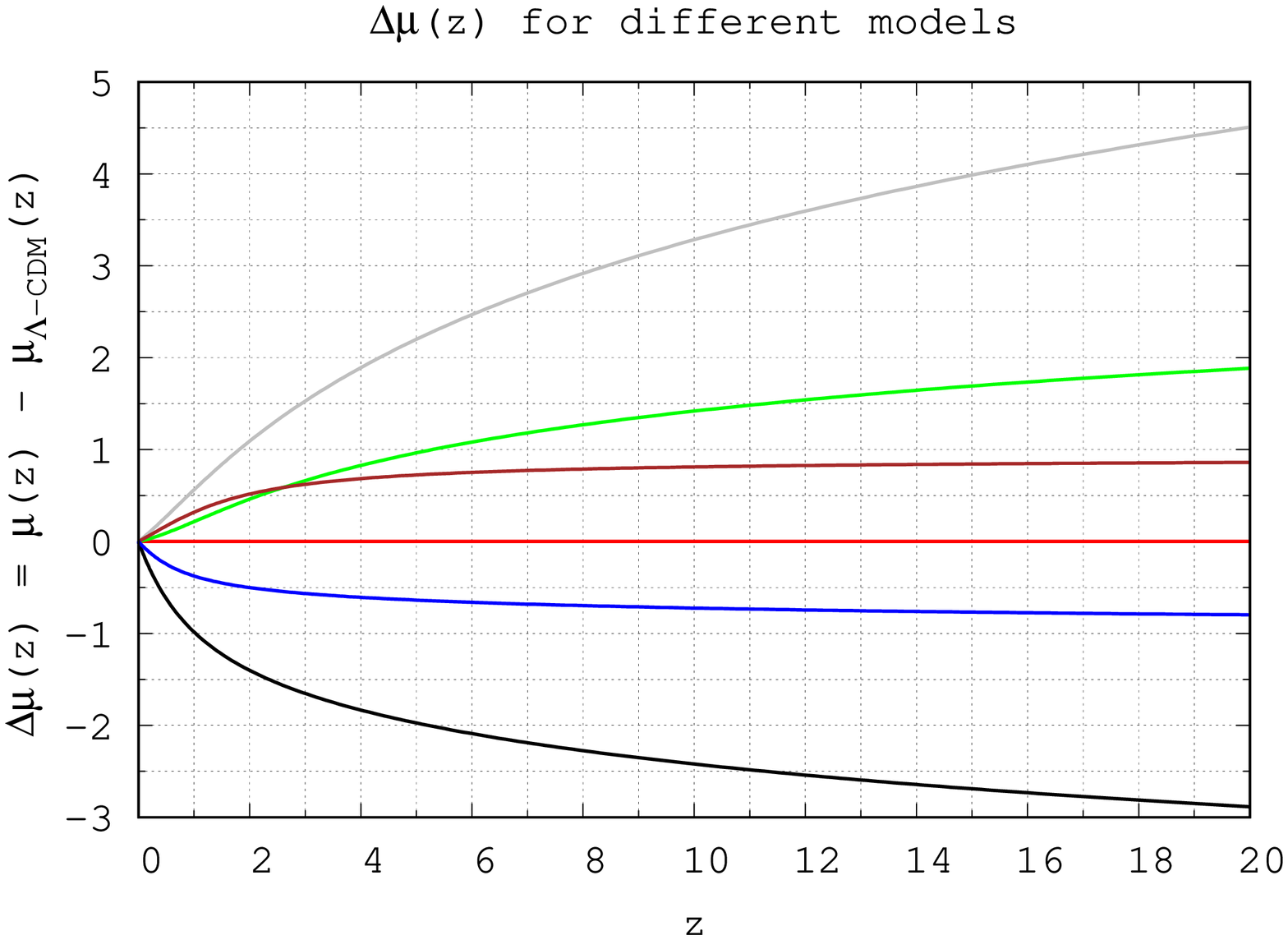} 
    \caption{The luminosity distance modulus  $\mu_L(z)$ 
    for six considered cosmological models described in Sec.\ref{standard-model-1} and Sec.\ref{models-param}. Left: direct distance modulus, Right: residuals from the standard $\Lambda$CDM model.
    }
    \label{fig:mu}
\end{figure*}

\subsubsection{The classical steady-state model}

The classical steady state (CSS) model can be used
for testing  the perfect cosmological principle, which asserts that the observable universe is basically the same at any time and at any place, so that the matter density in the expanding universe remains unchanged due to a continuous creation of matter~\citep{Hoyle2000}.
The HD for the CSS has specific behaviour that can be directly tested by observations.

Theoretical parameters of the CSS  model correspond
to those of the zero-curvature Friedmann model with an exponentially growing scale
factor $S(t) = S_0 e^{Ht}$, ($H=H_0$ = const). The metric distance is 
\begin{equation}
 r^\text{CSS}(z)= l(z) = R_{H}\, z 
 = r^\text{Vac}\,,
\end{equation}
where $R_{H}=c/H_0$, and it is the same as in the pure vacuum $\Lambda$CDM model.  

The luminosity distance modulus of the CSS model  coincides with the $\Omega_{\Lambda}= 1$ pure vacuum
$\Lambda$CDM model:
\begin{equation}
\label{mu-CSS}
\mu_L^\text{CSS} = 5 \log\,(R_{H}\,z) +\,
5\log (1 +z) + 25 + C_i(z) 
= \mu_L^\text{Vac} \,.  
\end{equation}

\subsubsection{The field-fractal model}

The field-fractal (FF) model can be used for testing  three basic principles of the SCM -- the gravity geometrization principle, the matter homogeneity   principle, and the space expansion paradigm.
The FF model is presented in~\citet{Baryshev2008} and \citet{Baryshev2012} and its modern status  allows one to formulate crucial observational tests, including the HD.

The FF model is based on the following basic assumptions:
(1) the gravitational interaction is described by the Poincare--Feynman field
gravitation theory in the flat Minkowski space-time, similar to all other fundamental physical interactions;
(2) the total distribution of matter (visible and dark) is described by a
fractal density law with the critical fractal dimension $D_\text{crit} = 2$; and (3) the cosmological redshift has global gravitational nature   within the fractal matter distribution.

An advantage of the FF model is that it solves the so-called Hubble--de Vaucouleurs paradox:  the coexistence of the observed strongly inhomogeneous distribution of visible matter of the local Universe ($0.0003<z<0.03$) and the linear Hubble Law ($z = H \,r/c$) on the same scales 
~\citep{Karachentsev2003,Baryshev2012}. 
In the expanding space of the SCM the Hubble Law is a strict mathematical consequence of the  homogeneous distribution of matter (ECP). So,  starting from very small scales 
$\sim 1$ Mpc there must be a homogeneous substance, which has the density 
$\rho_\text{dark} \gg \rho_\text{visible}$.

In the FF model the linear Hubble redshift--distance law is consistent with the strongly inhomogeneous distribution of matter on small scales $1<r<100$ Mpc.
Indeed, within a fractal galaxy distribution the global gravitational redshift of a source observed at a distance of $r$ will be  $z_\text{g} \propto \phi(r) / c^2 \propto M(r)/r \propto r^D/r \propto r^1$. Thus, it gives the linear redshift--distance law for the fractal dimension $D=2$. Taking into account the FF cosmological solution of the field equation for gravitational potential $\phi$, the redshift--distance relation is given by the following
expression in~\citet{Baryshev2008} and \citet{Baryshev2012}:
\begin{equation}
\label{z-grav-r}
    z_{g}(x) = \left( \frac{1}{2\sqrt{x}} I_1(4\sqrt{x}) \right)^\frac{1}{2}
    -1 \equiv W(x)\,,
\end{equation}
where $x=r/R_{H}$, $R_{H}=c/H_\text{g}$, and $I_1$ is the modified Bessel function.
The corresponding metric distance--redshift relation is \begin{equation}
\label{r-z-grav}
    r^\text{FF}(z) = R_{H} \,\, Y(z)\,,
\end{equation}
where $Y(z) = W^{-1} (z)$ is the inverse function of $W(z)$.
The properties of the gravitational redshift are analogous to those of the Doppler redshift and hence the luminosity distance modulus will be
\begin{equation}
\label{mu-FF}
\mu_L^\text{FF} = 5 \log\,(R_{H}\,Y(z)) +\,
5\log(1 +z) + 25 + C_i(z) \,.   
\end{equation}

\subsubsection{The tired-light model}

The tired-light (TL) model in the Euclidean static space~\citep[e.g.,][]{LaViolette1986} can be used as a toy model to demonstrate the importance of the time dilation cosmological effect. 
Within the framework of  the TL model, the cosmological redshift is caused by the photon energy $E=h\nu$ decrease proportional to the covered distance,  $h\nu_\text{obs} = Z\, h\nu_\text{emit}$, where $Z=e^{-\alpha r}$  is the Zwicky factor. Thus the distance--redshift relation is
\begin{equation}
\label{r-z-TL}
    r^\text{TL}(z) = R_{H} \, \ln{(1+z)} \,,
\end{equation}
where  $R_{H}=c/H_\text{0}$
and the luminosity distance modulus will be
\begin{equation}
\label{mu-TL}
\mu_L^\text{TL} = 5 \log\,(R_{H}\,\ln(1+z)) +\,
2.5\log(1 +z) + 25 + C_i(z) \,.   
\end{equation}


\subsubsection{Figures for considered models}
In the context of SCM we consider the  HD for three versions of the matter and dark energy density parameters: two $\Lambda$CDM models with $w=-1$ having $(\Omega_\text{m} =0.3; \,\,\Omega_\text{vac}=0.7;\,\, \Omega_{k}=0)$ and 
$(\Omega_\text{m} =0.1;\,\, \Omega_\text{vac}=0.9;\,\, \Omega_{k}=0 )$, and $w$CDM model with $w=-0.5$ having $(\Omega_\text{m} =0.5; \,\,\Omega_{w}=0.7;\,\, \Omega_{k} = 0.2)$. The metric and luminosity distances for these models are given by Eqs.~(\ref{r-z}), (\ref{l-z}), and (\ref{lum-dist-L}) and represented in Fig.~\ref{fig:dL}.
The luminosity distance modulus are given by Eq.~(\ref{mu-L-Frid}) and represented in Fig.~\ref{fig:mu}.

For all the considered models  the relations between luminosity and metric distances are given in Table~\ref{models}. The metric and luminosity distances as functions of redshift are represented in Fig.~\ref{fig:dL}. The luminosity distance modulus  for these models are represented in 
Fig.~\ref{fig:mu}.

\begin{table}\centering
\begin{tabular}{c|cccc} \hline
    Model               &   $r(z)/R_{H}$                  & $d_L(z)/r(z)$ & $d_E(z)/r(z)$                   \\
    \hline
    \emph{SCM} &   $\int 0^z\frac{dz'}{h(z')}$ & $(1+z)$       & $\sqrt{1+z}$               \\
    \emph{CSS}          &   $z$                         & $(1+z)$       & $\sqrt{1+z}$      \\
    \emph{FF}           &   $Y(z)$                      & $(1+z)$       & $\sqrt{1+z}$     \\
    \emph{TL}           &   $\ln{(1+z)}$                & $\sqrt{1+z}$  & $\sqrt{1+z}$  \\
\end{tabular}                     
\caption{
The dimensionless metric, luminosity and energy distances behaviour as functions of the redshift $z$ for the cosmological models described in Secs.~\ref{standard-model-1} and \ref{models-param}.
}\label{models}
\end{table}


\section{ LGRB as standard candles and Amati relation }
\label{LGRB standard candles and Amati relation}

\subsection{The observed LGRB parameters}

LGRB  sources are related to explosions of massive core-collapse SN in distant galaxies  \citep[e.g.,][]{Woosley1993, Woosley2006, Cano2017},
though up to now  there is no satisfactory theory of the  LGRB radiation's origin~\citep{Meszaros2014,Kumar2015,Peer2015,Willingale2017}. The observed LGRB photon spectrum is well approximated by the standard empirical model of~\citet{Band1993}
\begin{equation}
    N(E)\left[\frac{\text{photons}}{\text{keV~s~cm}^2}\right] = 
    \begin{cases}
        AE^\alpha e^{-E/E_0}, & E<(\alpha-\beta)E_0\\
        BE^\beta,             & E>(\alpha-\beta)E_0
    \end{cases},
    \label{bandfuction}
\end{equation}
where $N\,(E=h\nu /100$ keV) is the Band function or the usual differential photon spectrum (d$N$/d$E$), $\alpha, \beta$ are observed the low-energy and high-energy spectrum parameters, 
$E_0$ is the break energy, $A$ is the normalized factor, and
$B = A\,(\alpha-\beta)E_0\,\,e^{(\alpha-\beta)}$ is given from the smooth junction condition. 

An important LGRB parameter is the observed peak energy $E_\text{p}$ (frequency) of the  spectral energy distribution, i.e. the photon energy (frequency), at which the energy spectrum $\nu F_\nu = E^2N(E)$  is maximum. So, the rest-frame peak energy is
\begin{equation}
    \label{E-p-i}
    \qquad
 E_{\text{p},i} = E_\text{p} \, (1+z) \,.
\end{equation}

The bolometric fluence (erg~cm$^{-2}$)  is calculated from observed values by the equation
\begin{equation}
\label{S-bolo}
    S_\text{bolo}= F_E = 
    S_\text{obs} \frac{\int_\frac{1}{1+z}^\frac{10^4}{1+z} E\,N(E)\text{d}E} {\int_{E_\text{min}}^{E_\text{max}} E\,N(E)\text{d}E}\,,
\end{equation}
and the bolometric peak flux is
\begin{equation}
P_\text{bolo}= F_L = P_\text{obs} \frac{\int_\frac{1}{1+z}^\frac{10^4}{1+z} E\,N(E)\text{d}E} {\int_{E_\text{min}}^{E_\text{max}} N(E)\text{d}E}\,,
    \label{Pbolo}
\end{equation}
where $E_\text{min}$ and $E_\text{max}$ are the limits of the observed spectral energy range, and $P$ is in units [erg~cm$^{-2}$~s$^{-1}$]. 

From Eq.~(\ref{flux-E})  the isotropic radiated energy (erg) is given by the relation
\begin{equation}
    E_\text{iso} = 4\pi d_E^2 S_\text{bolo}  
    \,.
\end{equation}

From Eq.~(\ref{flux})  
the isotropic peak luminosity (erg~s$^{-1}$) is given by
\begin{equation}
    L_\text{p}=4\pi d_L^2 P_\text{bolo}\, .
\end{equation}

\subsection{Gravitational lensing and Malmquist biases}

\subsubsection{Observational distortion of theoretical LGRB parameters}

There are a number of inevitable  observational selection effects (e.g., limits on detector sensitivity, influence of the intervening matter, GL, and evolution), that potentially distort the measured  flux and fluence,  and hence the derived distance to a GRB. 
Thus the construction of the proper HD should take into account different selection and evolution effects,
which have been widely debated in the literature \citep[e.g.,][]{Schaefer2007,Liang2008,Wang2011b,Dainotti2013a,Dainotti2013b,Liu2015,Deng2016,Lin2016a,Lin2016b, Wang2016,WangG2017,Demianski2017a, Lloyd2019, Xue2019}.

However, a definite answer to the role of selection effects
in the LGRB luminosity estimation requires more data with known redshifts. The forthcoming THESEUS mission together with accompanying multimessenger observations  is expected to bring a solution to this fundamental question.

\subsubsection{Gravitational lensing along the LGRB line of sight}
\label{grav-lensing}
The apparent image of a distant source can be distorted by the GL effect of the mass density fluctuations (both visible and dark) along the line of sight. The GL of a variable source produces two main effects: it splits the source image and creates a time delay between different subimages of the source. 
According to modern observational surveys the GL plays an important role in cosmology. There is a sufficiently large probability for detection of strong lensing effects that allows one to study the total (dark and luminous) mass distribution of lenses  \citep{Ji2018,Acebron2020,CervantesCota2020,Shajib2019}.  For GL of high-redshift  LGRBs and quasi-stellar objects (QSOs), it is especially important to study protoclusters of low-luminosity star-forming galaxies, which were recently discovered at 
$z \approx 6.5$ \citep{Calvi2019}.

The next generation of cosmological surveys
will elucidate the value of weak and strong lensing biases in the true luminosity of distant sources and hence in observed magnitude--redshift relations at $z \sim 1$ \citep{Laureijs2011,Scolnic2019,CervantesCota2020}.

\emph{Basic physics of the GLB.}
Flux magnification of the split images due to GL should be taken into account in observations of distant compact sources, such as SNe, GRBs and QSOs. For the lens mass $M < 10^8$ $M_{\odot}$ the splitting angle is  $\theta < 0.3$ arcsec and a compact source will be observed as one image (for ordinary angular resolution). This gravitational magnification of the observed fluxes will lead to the GLB in the estimated luminosity of compact sources.

GL along a GRB line of sight magnify its flux due to different gravitating  structures  of the Universe, such as dark and luminous stellar mass objects, globular and dark stellar mass clusters, and galaxies and dark mass galaxy halos. 
Hence the GL can have a great impact on high-redshift LGRBs.

For estimation of the GLB  one must assume a GL model that includes three parts:
(1) a lensed object (source); (2) lensing objects (lenses); and (3) a distribution of lenses along the source line of sight. As a source, we  consider distant  SNe Ia, GRBs, and QSOs. 

The lensing object (gravitational lens) is defined by its mass density distribution. In accordance with the total  lens mass interval, usually one considers three types of lensing: microlensing ($M<10^3$ $M_{\odot}$), mesolensing 
($10^3 < M <10^8$ $M_{\odot}$) and macrolensing 
($M>10^8$ $M_{\odot}$). 
Gravitational lenses include both visible and dark matter objects. Examples are stars, stellar clusters, galaxies and galaxy clusters. The transparent gravitational lenses (like globular clusters and dwarf galaxies) have a larger cross-section for lensing effect due to the more complex structure of caustics. 

A very important part of the GL model is the assumed distribution of lenses along the line of sight. It plays a crucial role in  the magnification probability and hence in the GLB. 
The strongly inhomogeneous distribution of matter along the line of sight leads to a larger lensing probability than the usually considered homogeneous distribution of matter 
(due to lens clustering). Observational data on the large-scale structure of the local and distant Universe  reveal inhomogeneous (fractal-like) visible distribution of matter within scales larger than hundreds of megaparsecs \citep{Gabrielli2005,Baryshev2012,Courtois2013,Einasto2016,Lietzen2016,Shirokov2016,Tekhanovich2016}.

Gravitational mesolensing of QSOs by globular clusters was considered by~\citet{Baryshev1997},
taking into account the fractal distribution of matter along the line of sight.
They demonstrated that the King-type transparent lenses (which have additional conic caustics) together with fractal large-scale distribution of the lenses along the line of sight (which enhance the total cross-section) will essentially increase the lensing probability.

Another evidence for mesolensing was found by
\citet{Kurt2000} and \citet{Ougolnikov2001} 
in the Burst and Transient Source Experiment (BATSE) catalogue, where there are several GRB candidates that are probably lensed by intergalactic globular clusters.

\emph{Weak and strong GL.}
There are many papers considering the weak and strong gravitational lensing effect for SN Ia, LGRB and QSO observations
\citep{Holz1998b,Wang2002,Holz2005,Jonsson2008,  Jonsson2010,Wang2011b,Smith2014,Ji2018,Scolnic2019,CervantesCota2020}.

The light from distant compact sources is affected by the GL induced by inhomogeneous massive structures (visible and dark)  of the Universe. Since photons are conserved by the lensing, the mean flux over the sources is preserved. The observed flux can be magnified (or reduced) by the GL produced  by random mass fluctuations in the intervening matter distribution. This effect leads to an additional dispersion in GRB brightness.
The probability distribution function (PDF) of GL magnification has higher dispersion than a Gaussian distribution  
\citep{Valageas2000, Wang2002, Oguri2006}. 

A statistical approach for taking into account weak GLB and MB in the LGRB observations was suggested by \citet{Schaefer2007}. It  was developed by~\citet{Wang2011b}, where a sample of 116 LGRBs was analysed. They found that weak GLB  effect shifts the estimation of matter density  to lower values.
So  $\Omega_\text{m} $  shifts from 0.30 to 0.26
and the corresponding vacuum density $\Omega_{\Lambda}$ shifts from 0.70 to 0.74.

However, as emphasized in 
\citet{Scolnic2019} and \citet{CervantesCota2020}, one of the most important problems of the next generation cosmological measurements is
the theoretical uncertainty in the expected lensing magnification bias.  It is still one of the largest unknown systematic effects, as the lensing probability is sensitive to both large- and small-scale distribution of matter that is difficult to model. The forthcoming space missions, such as Euclid, will allow one to estimate  the value of weak and strong lensing biases in the magnitude--redshift relations at
$z \sim  1$ \citep{Laureijs2011} and beyond
\citep{Calvi2019}.

In particular, the mesolensing  and strongly inhomogeneous line of sight distribution of matter can lead to essential  changes in the estimated cosmological density parameters. We started the observational  program for optical study of the LGRB line-of-sight distribution of lensing galaxies using photometric and spectral galaxy redshifts
\citep[e.g.,][]{CastroTirado2018, SokolovJr2018,  Sokolov2018}.
Observations of deep fields in the directions of LGRBs reveal  galaxy clusters along the LGRB line of sights. 
This  program is important for estimation of GLB in LGRB catalogues.

\emph{Malmquist bias (MB).}
The GL of LGRBs produces  apparent increase of flux $P_\text{bolo}$ and fluence $S_\text{bolo}$ due to the gravitational lens magnification, which does not change frequency (i.e., the peak energy $E_\text{p}=h\nu_\text{p}$) of the lensed radiation. This can be misinterpreted as an evolution of the GRB luminosity.

If we take into account that there is a threshold for the detection in the burst apparent brightness, then, with GL, bursts just below this threshold might be magnified in brightness and detected, whereas bursts just beyond this threshold might be reduced in brightness and excluded. This observational selection effect is known as the Malmquist bias (MB) that plays an important role in observational cosmology
\citep{Baryshev2012}.

~\citet{Schaefer2007}  considered the GLB and the MB effects for a sample of 69 LGRBs. He found that the GLB and MB are smaller than the intrinsic error bars. However, as we emphasized above, modern observations reveal strong matter clustering
and so more complex lensing models must be studied further. Note that for THESEUS observations the MB will be shifted to larger distances due to better sensitivity.

\emph{Phenomenological model for GLB and MB.}
 The crucially important fundamental question on the role of the GLB in high-redshift 
SN Ia, LGRB and QSO data is still open and needs additional observational and theoretical studies.
Detailed observational study of the GLB is one of the primary tasks of future ground-based and space missions. However, already now, we can estimate quantitatively the value of combined contribution of the weak and strong GL effect and the MB, if we introduce  simple parametrization of the observed radiation flux from a distant source.

Let one consider a phenomenological approach
to the study of LGRB GL and  bias. It can be considered 
as the first step in taking into account the common action of the GLB and MB, together with the strongly inhomogeneous distribution of lenses along the line of sight. 
Consider a general bolometric fluence correction in the power-law form
\begin{equation}
\qquad
    S_\text{bolo}^\text{cor}=
    \frac{S_\text{bolo}} 
    {(1 + z)^{k}} \, ,
    \label{Scorr}
\end{equation}
where $S_\text{bolo}$ is the observed bolometric fluence,
$\text{k}$ is the corresponding bias parameter, which parametrizes  the strength of the apparent fluence magnification. According to Eq.~(\ref{Scorr})
the true value of GRB fluence 
$S^\text{true}_\text{bolo} =S^\text{cor}_\text{bolo} $ and it must be used to compare theory with observations.

Note, that according to \citet{Deng2016} there is an observed evolution of the intrinsic peak luminosity of the high-redshift LGRBs in the form  
$L_\text{p} \propto (1 + z)^{k_\text{p}}$ with $k_\text{p}=1.49 \pm 0.19$.
It is clear that at least a part of this ``evolution'' can be caused by the GLB$+$MB selection effects. 

As a testing value of the GLB$+$MB parameter, we consider $k=0.25,\,0.5$, and $0.75$ in Eq.~(\ref{Scorr}), which is less than  the observed luminosity increase  power-law exponent  $k_\text{p}=1.49$  in~\citet{Deng2016}. 

The solution of the fundamental problem (derivation of the
true value of the GLB and MB parameter $k$) will be based on  future
combined development of both lensing models (magnification PDF) and observations of inhomogeneous distribution of lenses along the LGRB line of sight. However now, using our phenomenological model, we can estimate the restrictions that follows from GLB and MB on the LGRB magnitude--redshift relation. Then we can compare the theoretical cosmological model predictions with observational data corrected by the bias.

\subsection{Amati relation}

LGRBs can be used as standard candles due to the relation discovered for GBRs from BeppoSAX observations. 
The correlation between the observed photon energy (frequency) of the peak spectral flux $E_{\text{p},i}$, which corresponds to the peak in the $\nu F_\nu$ spectra, and the isotropic equivalent radiated energy $E_\text{iso}$ was discovered and studied by~\citet{Amati2002, Amati2008} and \citet{Amati2013}.
The Amati relation can be written as
\begin{equation}
    \log{\frac{ E_\text{iso}^\text{A} }{ \text{1 erg} }}=a \log{\frac{ E_{\text{p},i} }
    {\text{300 keV} }} + b\,,
    \label{amati-law}
\end{equation}
where $E_{\text{p},i}$ is the GRB source rest-frame  spectrum peak energy given by Eq.~(\ref{E-p-i}),  ``$a$'' and ``$b$'' are Amati parameters.

The Amati correlation can be established for a sample of GRBs where redshifts $z$ are measured.
Additionally, the Amati coefficients can be calibrated by GRBs in the same range, where we have good statistic of SNe Ia and where we approximate the $d_L(z)$ -- luminosity distance as a function of redshift. 

\subsection{Extended Amati relation}

The Amati relation (Eq.~\ref{amati-law}) can be transformed into an extended Amati relation (the Amati law of GRB distances) by using Eqs.~(\ref{flux}), (\ref{flux-E}), (\ref{lum-dist-L}), and (\ref{lum-dist-E}) so the GRB luminosity distance is given by the equation
\begin{equation}
    d_L^\text{A}\,[\text{cm}] = \left( \frac{(1+z)E_\text{iso}^\text{A}(E_{\text{p},i})}{4\pi S_\text{bolo}} \right)^{\frac{1}{2}}\,\,. 
    \label{f:d_L}
\end{equation}
\noindent Taking into account Eq.~(\ref{Scorr}) we get the GRB luminosity distance modulus in the form
\begin{equation}
  \mu^\text{A}_\text{GRB} = 25 + \frac{5}{2} \left[ 
\log{ 
\frac{ (z+1)^{k+1} }{ 4\pi S_\text{bolo} } + 
a \log{ E_{\text{p},i} } + b 
} \right]\,, 
    \label{f:mu_final}
\end{equation}
where $a,\,b$ are the Amati coefficients, and $\text{k}$ is the  parameter of the GLB and MB. 
Hence the correction of the joint distance modulus due to action of GLB and MB is 
\begin{equation}
\Delta \mu_\text{eff} = 2.5 \log{ (z+1)^{k} }\,, 
    \label{f:dmu_GLMB}
\end{equation}
where the parameter $\text{k}$ is a measure of the observed HD distortion by the GLB and MB effects.

\section{SN Model-Independent Hubble Diagram}
\label{SN Model Independent Hubble Diagram}

\begin{figure*}
    \includegraphics[height=0.49\textwidth,angle=-90]{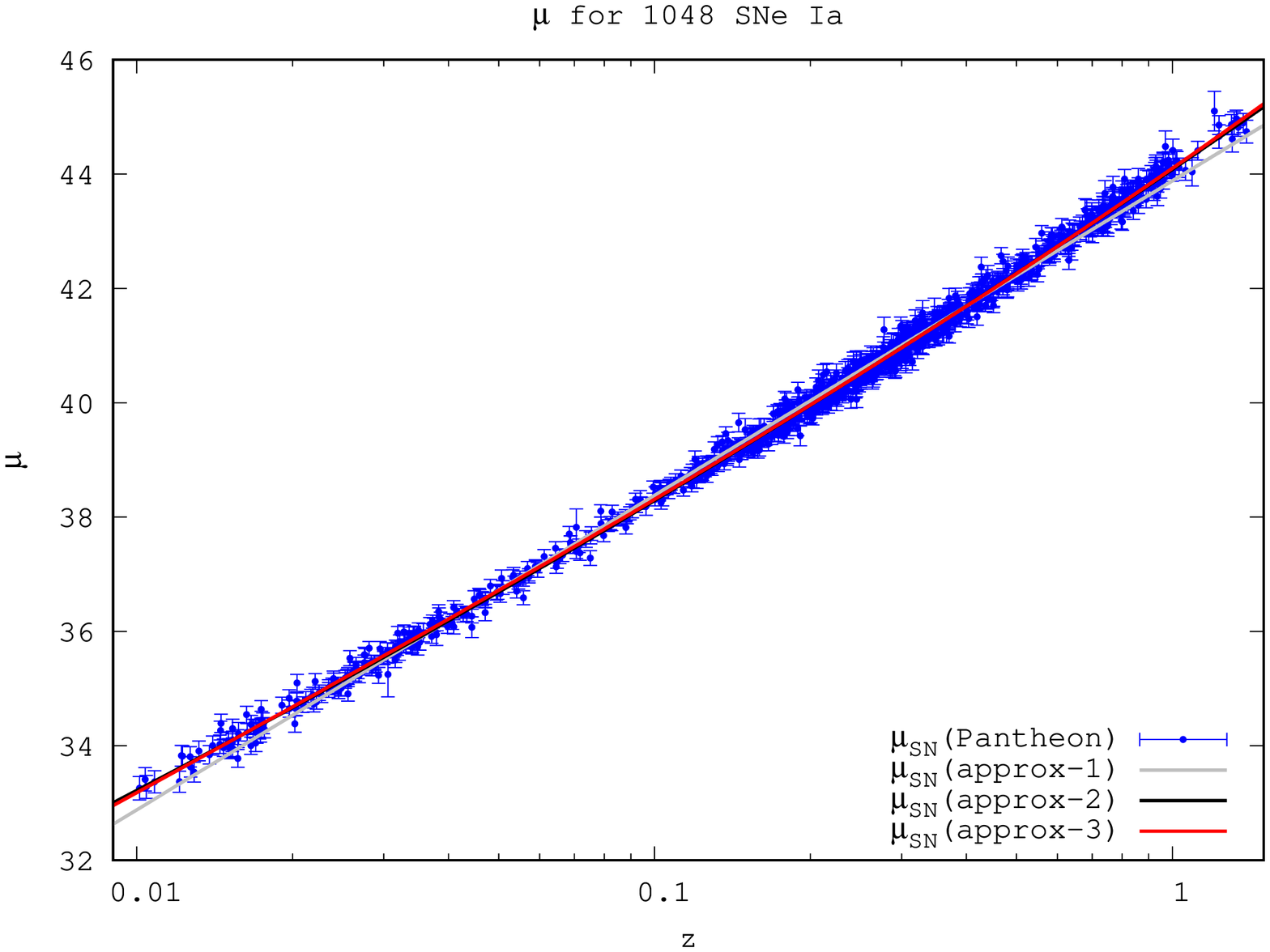}
    \includegraphics[height=0.49\textwidth,angle=-90]{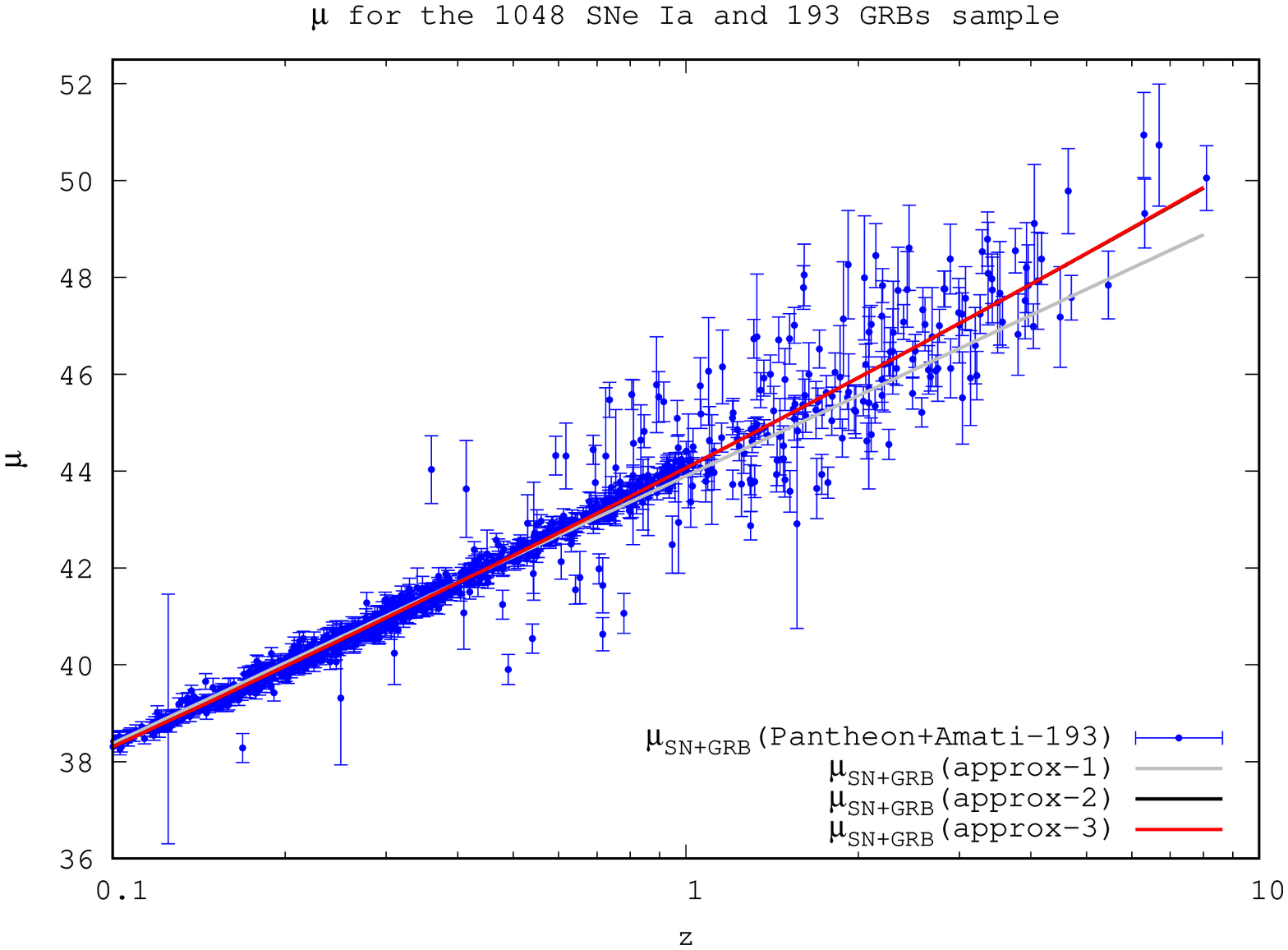}    
    \caption{The model-independent Hubble diagram (HD) for Pantheon SNe Ia (left) and for combined SNe Ia$+$LGRB (right) and its polynomial approximations. 
    }
    \label{fig:SN}
\end{figure*}

The SNe Ia model-independent HD can be directly inferred from the Pantheon observational data
\citep{Scolnic2018}, including 1\,048 SNe Ia. It is shown in Fig.~\ref{fig:SN} (left).
The luminosity distance values $d_L$ and the luminosity distance modulus  $\mu=m-M$ are given  by the relation 
\begin{equation}
    \mu_L^\text{SN} = 5\log{\frac{d_L^\text{SN}(z)}{\text{Mpc}}}+25\,,
    \label{mu}
\end{equation}
where the luminosity distance is expressed in megaparsecs for the given $z$.

If the parameters are determined using some cosmological model with adjusted fixed parameters, giving the dependence $d_L(z)$, then the HD obtained for GRBs cannot be correctly used to determine both the cosmological model and its parameters, since this approach involves the circularity problem~\citep[e.g.,][]{Kodama2008}.

Instead, the task is to determine the dependence $d_L(z)$ for GRBs without making assumptions about cosmology
(at least in the first approximation). To construct the  $d_L(z_i)$  we use the Pantheon observational data on $\mu(z_i)$ with errors in determining the distance modulus and effects of deviations from the standard candles. Index $i$ is indicated in the process of discreteness. 
The continuous function $d_L(z)$ must be smooth and monotonically increasing.

Without resorting to cosmology assumptions, we can approximate the dependence $d_L(z)$ in the logarithmic coordinates by some elementary mathematical function by the least-square method. 
Since the dependence $\mu(\log{z})$ is linear at small $z$, it is reasonable to approximate it with polynomials of small degree $n$, i.e.
\begin{equation}
    \mu(x) \approx P_n(x) = \sum_{i=0}^n p_i x^i\,,
    \label{Happrox}
\end{equation}
where $x=\log{z}$.

The function is linear for parameters, hence it is possible to apply the matrix least-squares method. 
In turn, the approximation of the dependence $d_L(z)$ will be exponential to the polylogarithmic function
\begin{equation}
    d_L(z)\,[\text{Mpc}] \approx 10^{0.2 P_n(x)-5}\,.
    \label{dL_approx_form}
\end{equation}

In this approach, we use the weighted least-squares method to minimize
\begin{equation}
    \chi^2 = \sum_{i=1}^N w_i (y_i - P_n(x_i))^2\,,
    \label{x2}
\end{equation}
where $x_i=\log z_i$, $y_i=\mu_i$, and $w_i=1/\sigma^2_{\mu_i}$ for each $i$ of $N$ SNe Ia by calculating the coefficients of the polynomial $P_n$ and their errors.

Calculations were made for polynomials of degree $1-3$, the results are shown in Table~\ref{tab:SNptab}.  Approximations of Eq.~(\ref{Happrox}) are shown in Fig.~\ref{fig:SN}.
The linear approximation is a good fit for small redshifts and the differences between polynomials of degrees $2$ and $3$ are small. 

\begin{table*}\centering
\begin{tabular}{c||c|c|c|c|c|c} \hline
Sample & Degree             & $p_0$            & $p_1$           & $p_2$            & $p_3$ \\ 

\hline

& 1 & $43.88 \pm 0.02$ & $5.50 \pm 0.01$ &  --  &   --                        \\ 
Pantheon & 2 & $44.08 \pm 0.01$ & $6.15 \pm 0.03$ &  $0.36 \pm 0.02$ &   --                         \\ 
& 3 & $44.10 \pm 0.01$ & $6.28 \pm 0.07$ &  $0.57 \pm 0.09$ &  $0.08 \pm 0.03$              \\ 

\hline

& {1} & $43.896 \pm 0.000$	 & $5.520 \pm 0.000$ &  --  &   --	 \\ 
SNe+LGRBs & {2} &  $44.062 \pm 0.000$	 & $6.096 \pm 0.002$	 & $0.341 \pm 0.001$  &  --	 \\ 
 & {3} & $44.063 \pm 0.000$	 & $6.099 \pm 0.005$	 & $0.347 \pm 0.011$	 & $0.003 \pm 0.002$	 \\ 

\hline

\end{tabular}                     
\caption{
The best approximation coefficients (Eq.~\ref{Happrox}) of the luminosity distance modulus $\mu(\log{z})$ for the Pantheon SNe Ia ($z_\text{cmb}$) and LGRB samples.
}\label{tab:SNptab}
\end{table*}

The polynomials  can be used for the junction condition between SNe Ia and LGRB data in the interval $0.01 <z<1.0$, where cosmological and selection effects are small. In  Fig.~\ref{fig:SN},  we also show the polynomial approximation of our total sample of SNe Ia and LGRBs for the total redshift interval $0.1 <z<10$.

\section{LGRB Hubble Diagram for different cosmological models }
\label{LGRB Hubble Diagram for different cosmological models}

The cosmological models considered in Sec.\ref{Hubble Diagram in Cosmological Models} cover a wide interval of deflections from the standard $\Lambda$CDM theoretical predictions for the luminosity distance modulus. Now we take the LGRB data to perform the high-redshift test of these models.

\subsection{LGRB sample}

We use the LGRBs sample of~\citet{Amati2019} for which there is a table of calculated luminosity distance moduli for total redshift interval.

As we emphasized above,  the Hubble law of linear distance--redshift relation starts immediately beyond the border of the Local Group~\citep{Karachentsev2003}. However there are a small number of LGRBs at redshifts $z<1$. This is why we have to use the SNe Ia observations, which give reliable luminosity distance--redshift data up to redshifts $\sim1$. The LGRB data that have a common region of redshifts with the SN data must obey the junction condition, which is fulfilled for our sample.

\subsection{Results for different assumed basic cosmological relations}
\label{Comparison}

\begin{figure*}
    \includegraphics
        [height=0.49\textwidth,angle=-90]
        {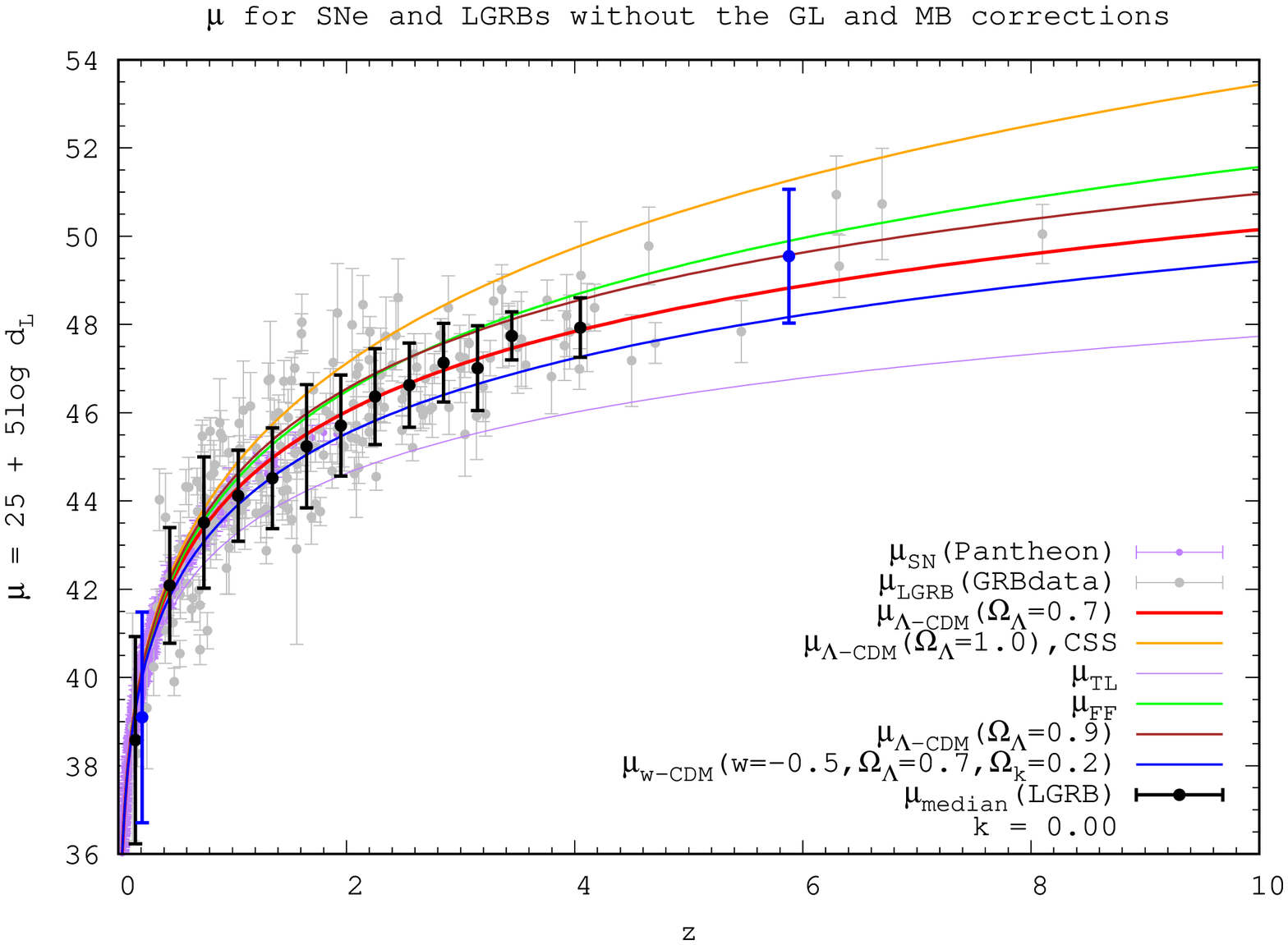}
    \hfill
    \includegraphics
        [height=0.49\textwidth,angle=-90]
        {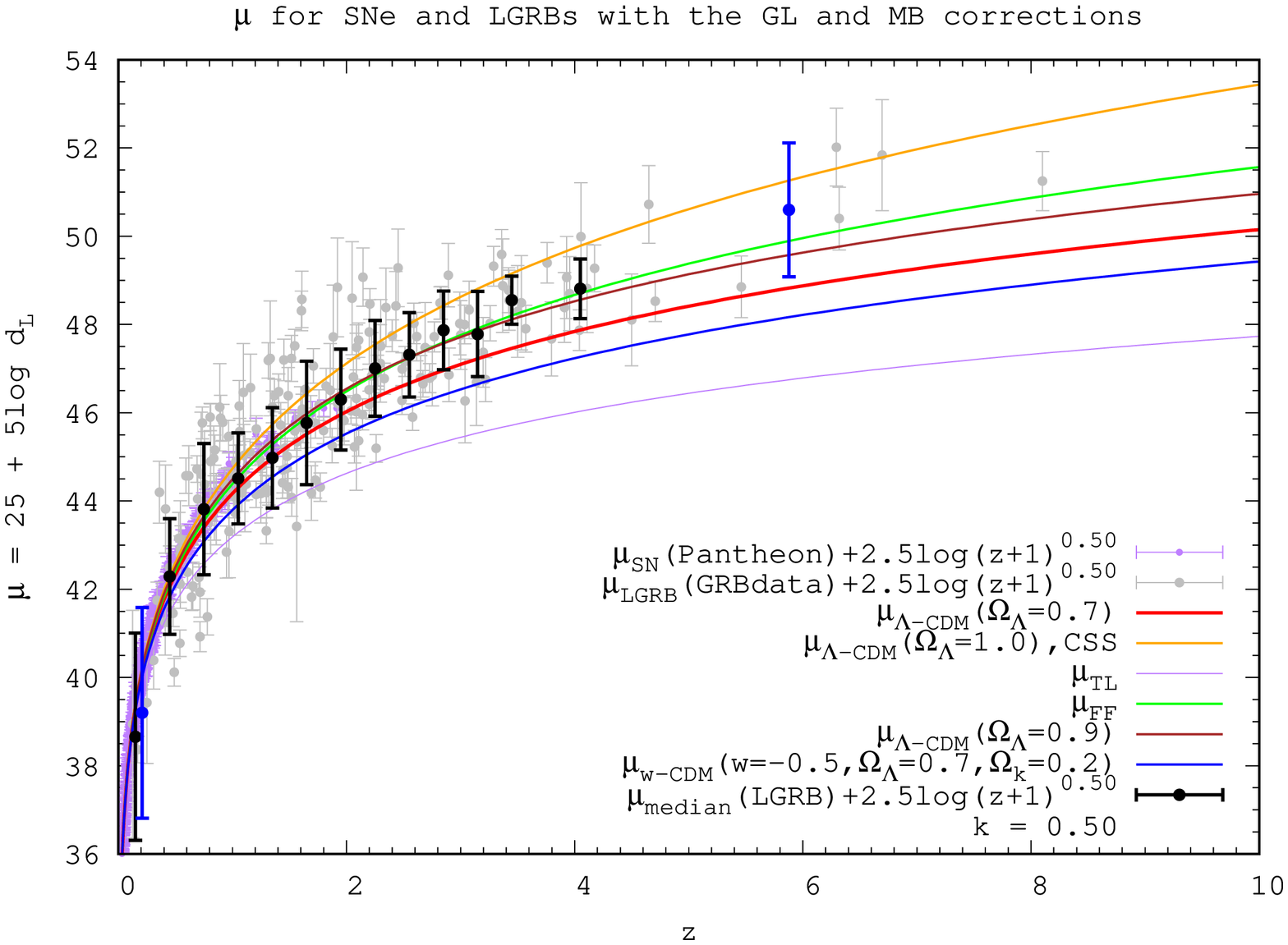}
    \vfill         
    \includegraphics
        [height=0.49\textwidth,angle=-90]
        {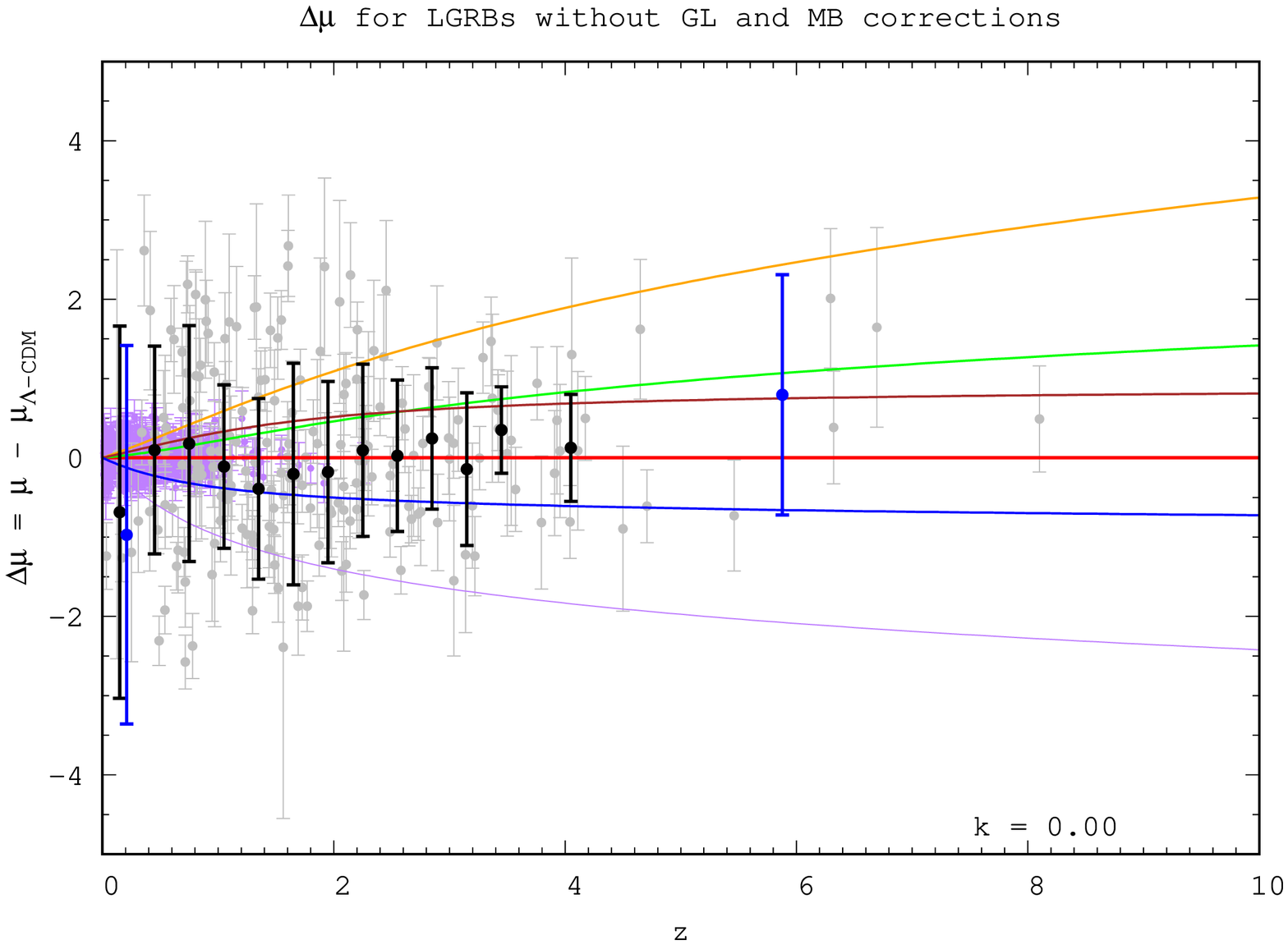}
    \hfill
    \includegraphics
        [height=0.49\textwidth,angle=-90]
        {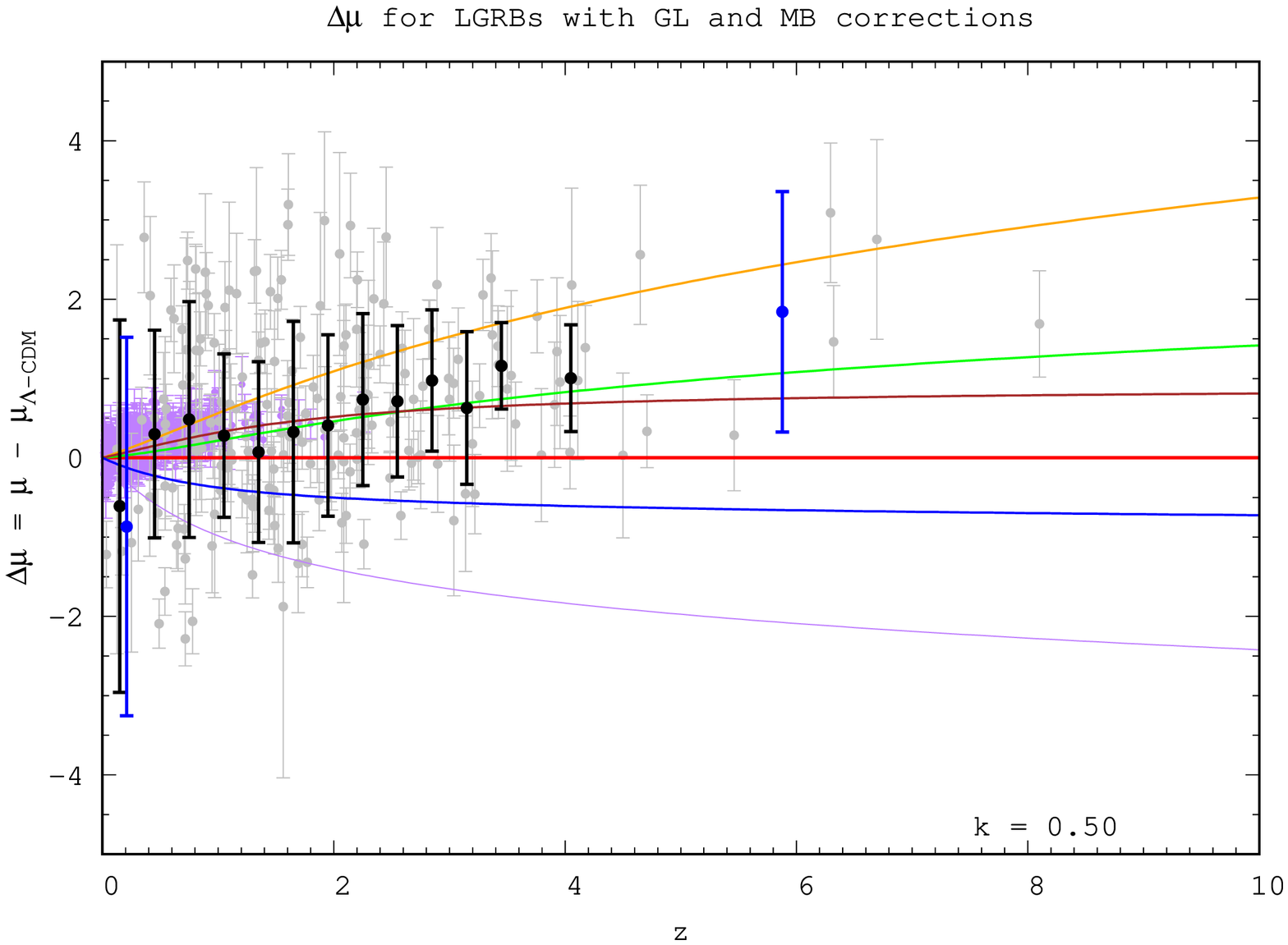}
        \caption{ 
        \emph{Top panels}: the observed luminosity distance modulus $\mu$ versus redshift (HD) in linear $z$-scale for the SN Ia and LGRB samples. Black points are the  median values of $\mu$ with linear step $\Delta z=0.3$ for the LGRB sample.
        \emph{Bottom panels}: the residuals $\Delta \mu$ from the standard $\Lambda$CDM model  for the observed luminosity distance modulus. \emph{Left}: without the correction for the GLB and MB. \emph{Right}: corrected with $k=0.5$.
        The colour curves correspond to the   cosmological models defined in Section~\ref{Hubble Diagram in Cosmological Models}.  
    }   
    \label{fig6:HD193lin}
\end{figure*}

In Figure~\ref{fig6:HD193lin} (and in Fig.~\ref{fig7:HD193log}), we present the LGRB HDs for the total 1\,048 SN Ia plus 193 LGRBs of the~\citet{Amati2019} sample. Theoretical models were calculated for the flat 
$\Lambda$CDM with 
$\Omega_{\Lambda}=0.7,\,0.9$, and $1.0$; the $w$CDM with a positive curvature having
$w=-0.5$,  $\Omega_{\Lambda}=0.7$, and $\Omega_{k}=0.2$. Different basic cosmological assumptions are illustrated by calculations of the luminosity distance modulus for CSS, FF and TL models (Sec.~\ref{Hubble Diagram in Cosmological Models}). The Hubble parameter is $h = 0.7$ for all models.

Figure~\ref{fig6:HD193lin} demonstrates the SNe Ia and LGRB  data in the linear redshift scale without taking into account the GLB and MB ($k=0$ in Eq.~\ref{Scorr}) and an example with the parameter $k=0.5$, which takes into account the possible GLB and MB. The deviations of the luminosity distance moduli and of the linear scale median values ($\Delta z=0.3$) of the LGRB sample from the standard $\Lambda$CDM model are also shown. Figure~\ref{fig7:HD193log} shows the same data but in logarithmic scale and with the logarithmic scale median values ($\Delta \log{z}=0.1$) of the LGRB sample. The blue points mean median values at $z<0.35$ and $4.2<z$, respectively. The legend in all plots is the same.

\begin{figure}
    \includegraphics
        [height=0.49\textwidth,angle=-90]
        {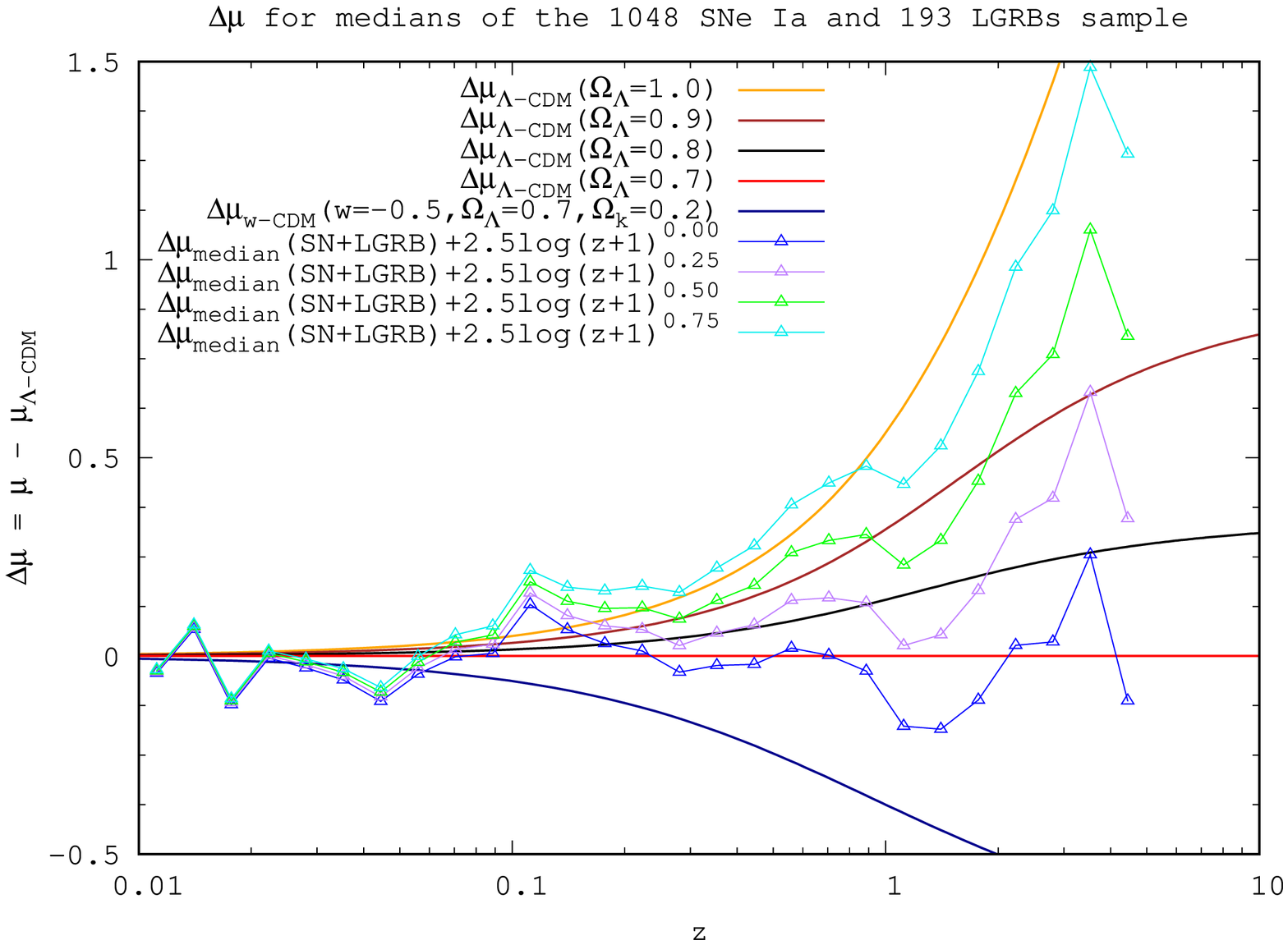}
    \caption{The residuals $\Delta \mu$  from the standard $\Lambda$CDM model   for the observed luminosity distance modulus $\mu$ calculated for different values of the  GLB and MB parameter $k=0.00,\,0.25,\,0.50$, and $0.75$. The colour curves correspond to the standard cosmological models (SCMs) with $\Omega_{\Lambda} = 0.7,\,0.8,\,0.9$, and $1.0$ and positive curvature $w$CDM model with $w=-0.5$, $\Omega_\Lambda =0.7$, and $\Omega_{k} =0.2$   defined in Section~\ref{Hubble Diagram in Cosmological Models}.  
    }
    \label{fig8:HD}
\end{figure}


To illustrate different values of the parameter $k$,  in Figure~\ref{fig8:HD},  
we present the results in residual form for the cases $k=0.0,\,0.25,\,0.50$, and $0.75$.
We also consider three simple statistical measures of the $\chi^2$ test in form from~\citet[p. 13]{Pearson1957}.
The first statistic uses  all LGRB sample points, 

\begin{equation}
    \chi^2_1 = \sum_{i=1}^{N_\text{GRB}} 
        \frac
            { (\mu_i - \mu_\text{model})^2 }
            { \mu_\text{model} }\,,
    \label{X2_1}
\end{equation}
where $N_\text{GRB}$ is the GRB number in each sample, $\mu_i$ is the GRB distance modulus, and $\mu_\text{model}$ is the considered  cosmological model distance modulus at the same redshift $z_i$. 

The second one uses the uniform linear median values for bins,
\begin{equation}
    \chi^2_2 = \sum_{j=1}^{N_\text{bins}} 
        \frac
            { (\mu_j - \mu_\text{model})^2 }
            { \mu_\text{model} }\,,
    \label{X2_2}
\end{equation}
where $N_\text{bins}$ is the number of bins in each sample, $\mu_j$ is the median distance modulus, and $\mu_\text{model}$ is the cosmological model distance modulus at the same redshift $z_j$.

The third one uses the uniform logarithmic scale median values for bins,
\begin{equation}
    \chi^2_3 = \sum_{k=1}^{N_\text{bins}} 
        \frac
            { (\mu_k - \mu_\text{model})^2 }
            { \mu_\text{model} }\,,
    \label{X2_3}
\end{equation}
where $N_\text{bins}$ is the number of bins in each sample, $\mu_k$ is the uniform in logarithmic scale median distance modulus, and $\mu_\text{model}$ is the cosmological model distance modulus at the same redshift $z_k$.

The results of our calculations of the statistics for the original and corrected (for GL and MB effects) LGRB sample of~\citet{Amati2019} are represented in Table~\ref{tab:X2}. The detailed description of the models is given in Sec.~\ref{Hubble Diagram in Cosmological Models} and Sec.~\ref{LGRB standard candles and Amati relation}.

From Table~\ref{tab:X2} we can see that uncorrected LGRB data correspond better to the standard parameters of the $\Lambda$CDM model (see the column for $\Omega_{\Lambda}=0.7$). Whereas the GLB and MB shift the fit towards the $\Lambda$CDM model with $\Omega_{\Lambda}=0.9$ for $k=0.5$.
Among alternative cosmological assumptions the corrected LGRB data are consistent with the field-fractal model (column FF), but reject the tired-light model (column TL) .

\begin{table*}\centering
    \begin{tabular}{cccccccc} \hline
        $\sqrt{\chi^2}$ & $k$ & $\Lambda$CDM($\Omega_\Lambda=1.0$) & FF & $\Lambda$CDM($\Omega_\Lambda=0.9$) & $\Lambda$CDM($\Omega_\Lambda=0.7$) & $w$CDM($w=-0.5,~\Omega_\Lambda=0.7,~\Omega_{k}=0.2$) & TL \\ \hline
$\chi_1$ &  0.00 &  3.051 &  2.405 &  2.411 &  2.302 &  2.572 &  3.735 \\
$\chi_2$ &  0.00 &  0.611 &  0.289 &  0.295 &  0.152 &  0.293 &  0.742 \\
$\chi_3$ &  0.00 &  0.553 &  0.274 &  0.280 &  0.140 &  0.237 &  0.623 \\
\hline
$\chi_1$ &  0.25 &  2.702 &  2.290 &  2.295 &  2.425 &  2.884 &  4.219 \\
$\chi_2$ &  0.25 &  0.461 &  0.171 &  0.181 &  0.223 &  0.438 &  0.899 \\
$\chi_3$ &  0.25 &  0.424 &  0.166 &  0.175 &  0.180 &  0.359 &  0.759 \\
\hline
$\chi_1$ &  0.50 &  2.441 &  2.321 &  2.325 &  2.674 &  3.273 &  4.728 \\
$\chi_2$ &  0.50 &  0.317 &  0.153 &  0.165 &  0.354 &  0.589 &  1.057 \\
$\chi_3$ &  0.50 &  0.298 &  0.130 &  0.141 &  0.287 &  0.490 &  0.896 \\
\hline
$\chi_1$ &  0.75 &  2.298 &  2.491 &  2.494 &  3.018 &  3.716 &  5.255 \\
$\chi_2$ &  0.75 &  0.191 &  0.257 &  0.265 &  0.501 &  0.743 &  1.215 \\
$\chi_3$ &  0.75 &  0.186 &  0.207 &  0.214 &  0.412 &  0.623 &  1.034 \\
        \hline
    \end{tabular}                     
    \caption{The $\chi_1$, $\chi_2$, and $\chi_3$ statistics for the~LGRB sample fitting    by considered cosmological models (Section~\ref{Hubble Diagram in Cosmological Models}).}
    \label{tab:X2}
\end{table*}

\section{Discussion and conclusions}
\label{Discussion and conclusions}

Modern astrophysical facilities open new possibilities for testing the basis of the cosmological models by using the observational approach to cosmology developed by Hubble--Tolman--Sandage  in the 20th century.
In the beginning of the 21st 
century the multimessenger astronomy allows one to test the fundamental initial assumptions of cosmological models with a higher accuracy and in a very wide redshift interval up to $z \sim 10$, instead of $z\sim 0.003$ in the first cosmological observations.

The high-redshift HD  for LGRBs can be used as  a necessary 
condition for plausibility of  a cosmological model, because it can test the directly observed flux--distance--redshift relation. In Section \ref{Hubble Diagram in Cosmological Models} we considered several examples of the cosmological 
$\mu\, \text{vs} \, z$ relations (the HD), which correspond to specific initial assumptions in the  cosmological models.

Existing LGRB data demonstrate that the Amati relation can be used for construction of the observed HD at the model-independent level, hence it is useful for testing theoretical models. However, the main uncertainty when comparing the observational HD and the theoretically predicted $\mu \,\text{vs} \, z$  relation is the problem of observational data distortion by different kinds of evolution and selection effects.

As emphasized in \citet{Scolnic2019}, and \citet{CervantesCota2020}, one of the most important problems of the next generation cosmological measurements is
the theoretical uncertainty in the expected lensing magnification bias.  It is still one of the largest unknown systematic effects, as the lensing probability is sensitive to both large- and small-scale distribution of matter for which there is no analytical model.
So this crucially important fundamental question on the role of the GLB in high-redshift 
SN Ia, LGRB and QSO data is still open and needs more  observational and theoretical studies.

Thus, an important obstacle for derivation of true cosmological parameters from the high-redshift LGRB HD is to correctly account for the GL. Flux magnification  and  MB play a crucial role in comparison of different cosmological models. For example, ``observed'' evolution of the LGRB peak luminosity 
$L_\text{p} \propto (1 + z)^{k_\text{p}}$ \citep{Deng2016}
can be partly caused  by the GLB  and MB.

In a previous study of the LGRB weak lensing statistical effects \citep{Schaefer2007,Wang2011b} an analytical lensing model was developed for calculation of the distance dispersion from the universal probability distribution function.
Modern observational data reveal a very complex large-scale distribution of matter (dark and luminous), which is difficult to model \citep{Scolnic2019}, but which can 
essentially change the probability of the GL. Especially for the high-redshift LGRBs, there is an important additional contribution to the lensing bias from the recently discovered population of protoclusters of low-luminosity star-forming galaxies at redshifts $z\sim 6$ \citep{Calvi2019}.

In our paper, as the first step for a quantitative examination of the possible lensing effects, we introduce a phenomenological one-parameter $k$-model (Section~\ref{grav-lensing}). In this way one can estimate the total GL effect produced by 
a combination of weak and strong lensing together with the MB. Though such an approach does not allow one to study the relative contributions and redshift evolution of these parameters, it gives the possibility to get restrictions on the considered cosmological models.

We extend the \citet{Wang2011b}  findings of a small shift of $(\Omega_\text{m},\,\Omega_{\Lambda})$ by the weak GL.  Our simple lensing model phenomenologically   describes the total bias due to strong and weak GL together with the strong inhomogeneous distribution of lenses along the LGRB line of sight
(Section~\ref{grav-lensing}). 

According to our analysis of the sample of 193 LGRBs
\citep{Amati2019}, we conclude that the high-redshift HD, corrected for the GLB and MB effects by the parameter $k$, points to a tendency to more vacuum-dominated $\Lambda$CDM models:  for 
$k \rightarrow 0.5$ we get
$\Omega_\Lambda \rightarrow 0.9$ and
$\Omega_\text{m} \rightarrow 0.1$ 
(see Figs.~\ref{fig6:HD193lin} and~\ref{fig7:HD193log}).
It is interesting to note that our results show that the positive curvature $w$CDM model with $w=-0.5$, $\Omega_\Lambda =0.7$, and $\Omega_{k} =0.2$ does not pass the HD test for all values of the bias parameter $k$.

Existing LGRB observations in the redshift interval 1--10  allow one to get new  restrictions on several examples of alternative theoretical flux--distance--redshift relations considered in Sec.\ref{Hubble Diagram in Cosmological Models}.
We showed (see Fig.~\ref{fig8:HD} and Table~\ref{tab:X2}) that the high-redshift LGRB  HD, corrected for the GLB and MB effects, could be consistent with the CSS model  if
$k \rightarrow 0.75$. It is also compatible
with the FF cosmological model for $k \rightarrow 0.5$, but it
rejects the TL assumption for all values of the bias parameter $k$.

Derivation of the true value of the GL and MB parameter $k$ will be based on the future development of both lensing models 
and observations of the inhomogeneous 
distribution of lenses  along the LGRB line of sight.
The forthcoming space missions, such as Euclid, will elucidate the value of weak and strong lensing biases in the magnitude--redshift relations at $z\sim 1$
\citep{Laureijs2011}.

Future THESEUS space observations of GRBs and accompanying  multimessenger ground-based  studies, including large [in particular Gran Telescopio Canarias (GTC), Bolshoi Teleskop Alt-azimutalnyi (BTA), and Elbrus-2] and even 1-m class optical telescopes can be used as a powerful tool for testing the basic cosmological principles. 
In particular, we started the program for optical study of the LGRB line-of-sight distribution of lensing galaxies 
\citep{CastroTirado2018, Sokolov2018, SokolovJr2018}.

The THESEUS GRB mission  will provide several hundreds of LGRBs with measured redshifts and spectral peak energy $E_\text{p}$ and together with optical line-of-sight observations will allow one to take into account the GLB and MB. This opens new possibilities for using the LGRB HD for checking the basic assumptions of cosmology, a prerequisite for establishing the observationally based true world model.

\section*{Acknowledgements}
We thank the anonymous reviewer for important suggestions that helped us to improve the presentation of our results.
We are grateful to P. Teerikorpi, A. J. Castro-Tirado, C. Guidorzi, and D. I. Nagirner for useful discussions and comments.
The work was performed as part of the government contract of the SAO RAS approved by the Ministry of Science and Higher Education of the Russian Federation.  

\bibliography{references_A}



\appendix

\section{abbreviations}

\itemize{

\item HD    --    Hubble diagram                

\item SN(e)   -- supernova(e)                                    
\item GRB(s) --   gamma-ray burst(s)                      

\item LGRB(s) --  long gamma-ray Burst(s)             

\item FLRW      -- Friedmann--Lemaitre--Robertson--Walker              
\item SCM       -- standard cosmological model                  
\item THESEUS -- Transient High-Energy Sky and Early Universe Surveyor
(space mission)
\item Euclid  -- space mission

\item MB  --   Malmquist bias          
\item GL        -- gravitational lensing                            
\item GLB    --   gravitational lensing bias                   
\item GRT       -- general relativity theory                        
\item ECP  --  Einstein's cosmological principle            
\item EoS(s)   -- Equation of state(s)                            
\item CSS  --  classical steady-state model          
\item FF        -- field-fractal model   
\item TL        -- tired-light model   
\item QSO(s)   -- quasi-stellar object(s)
}

\section{Metric and luminosity distances}

\subsection{Metric distance--redshift relations }
There are two kinds of metric distance in the FLRW model.
The internal metric distance $r$ between the observer and the source at redshift $z$ at cosmic time $t=t_0$  in general expanding space models is given by the relation
\begin{equation}
  \label{r-z}
  r(t_0,z)=r(z)= \frac{c}{H_0}\int_0^z\frac{\text{d}z'}{h(z')}\,,
\end{equation}
and the external metric distance $l$ will be
\begin{equation}
  \label{l-z}
  l(t_0,z)=l(z)= S(t_0)I_{k}(r(z)/S(t_0))\,, 
\end{equation}
where the scale factor at present epoch
  $S(t_0)=S(0)= c/H_0 
  (k/\Omega_{k})^\text{1/2}$.
Here the normalized Hubble parameter $h(z)=H(z)/H_0$ in the case of the $w$CDM model, having two fluids with equations of state for cold matter $p=0$ and for quintessence (dark energy) $p=w\rho c^2$ ($w<0$), 
is given by Friedmann's equation as
\begin{equation}
    h(z)=\sqrt{\Omega^0_\text{m}(1+z)^3 +
    \Omega^0_\text{DE}(1+z)^\text{3(1+w)} -
    \Omega^0_{k}(1+z)^2}\,,
    \label{Hubble-w-CDM}
\end{equation}
where  density parameters are defined as $\Omega_i=\rho_i/\rho_\text{c}$ with the critical density 
$\rho_\text{c} = 3H^2/8\pi G$, 
and $w$ is the dark energy equation of state (EoS) parameter.

The curvature density parameter is $\Omega_{k}  = kc^2/S^2H^2 = (\Omega_\text{tot}-1)$. We use the definition of $\Omega_{k}$ that has the same sign as the curvature constant 
${k}$.  Note that in a number of papers they use another definition 
$\Bar{\Omega}_{k} = -\Omega_{k}$
so the negative curvature space has a positive 
bar curvature density parameter.

In the general case of several non-interacting fluids $\rho_i$ with EoSs $p_i/\rho_ic^2 =\alpha_i$, the total density parameter is $\Omega_\text{tot}= \Sigma_i \Omega^0_i
(1+z)^{3(1+\alpha_i)}$.
For $w=-1$ we have $p=-\rho$ and constant cosmological vacuum density:
$\Omega_\Lambda =\Omega^0_\text{DE}$ = const, which is called the $\Lambda$CDM model. 
For dark energy parameter $w < 0$ the model is called quintessence $w$CDM, and for $w<-1$ we have the ``phantom'' model.

\subsection{Luminosity and energy distances }

Consider a light source  at the metric distance $r(t_0,\chi)=S(t_0)\chi$ detected at the time $t=t_0$ in the Friedmann universe. 

The source emitted light at the time $t=t_1$  isotropically around it with the bolometric (total) luminosity $L_\text{iso}$ (erg\,s$^{-1}$).
Draw a sphere with the source in the centre and the observer at the surface at the moment of reception ($t =t_0$). 

\begin{figure*}

    \includegraphics[height=0.49\textwidth,angle=-90]{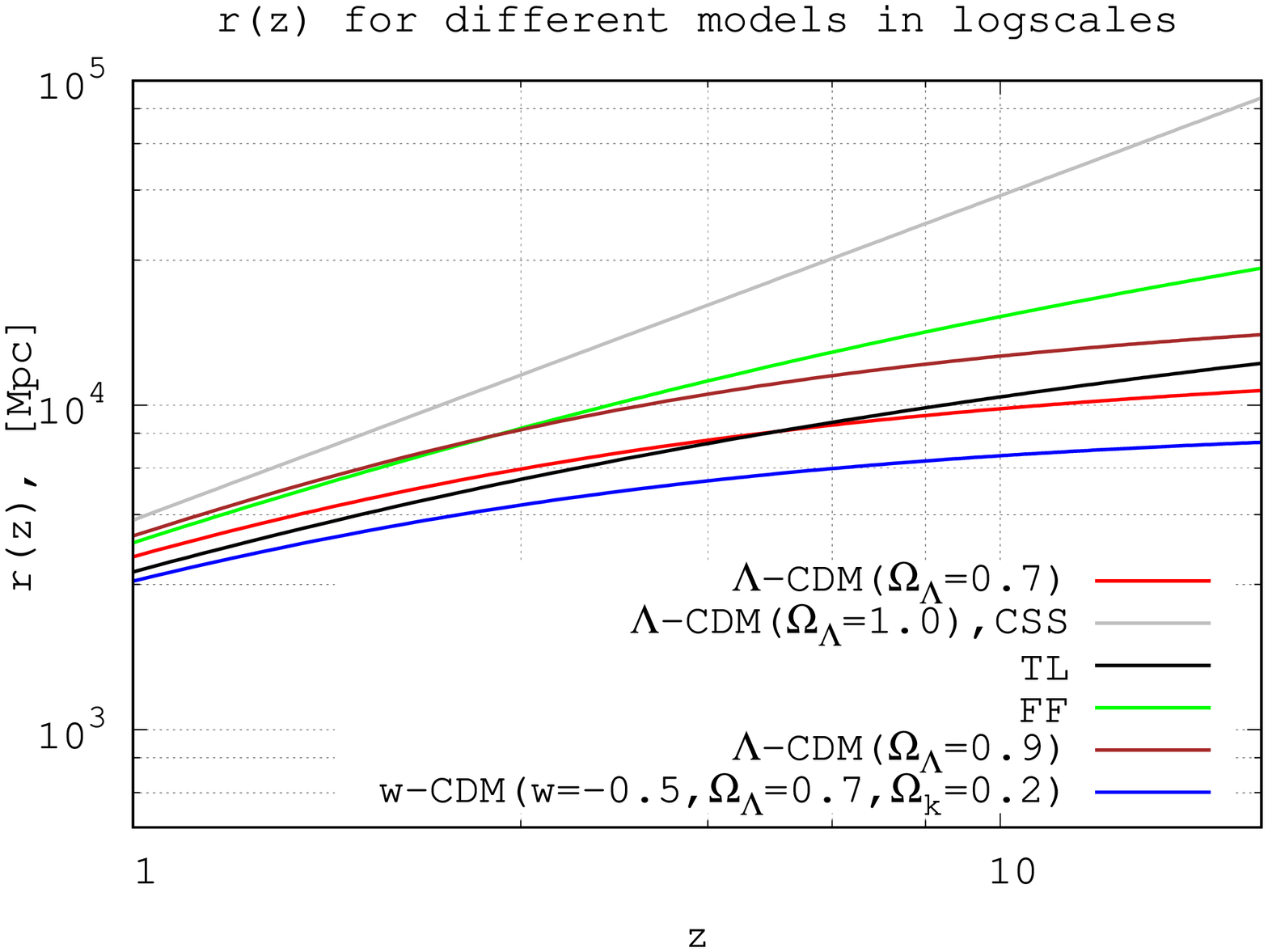}
    \hfill
    \includegraphics[height=0.49\textwidth,angle=-90]{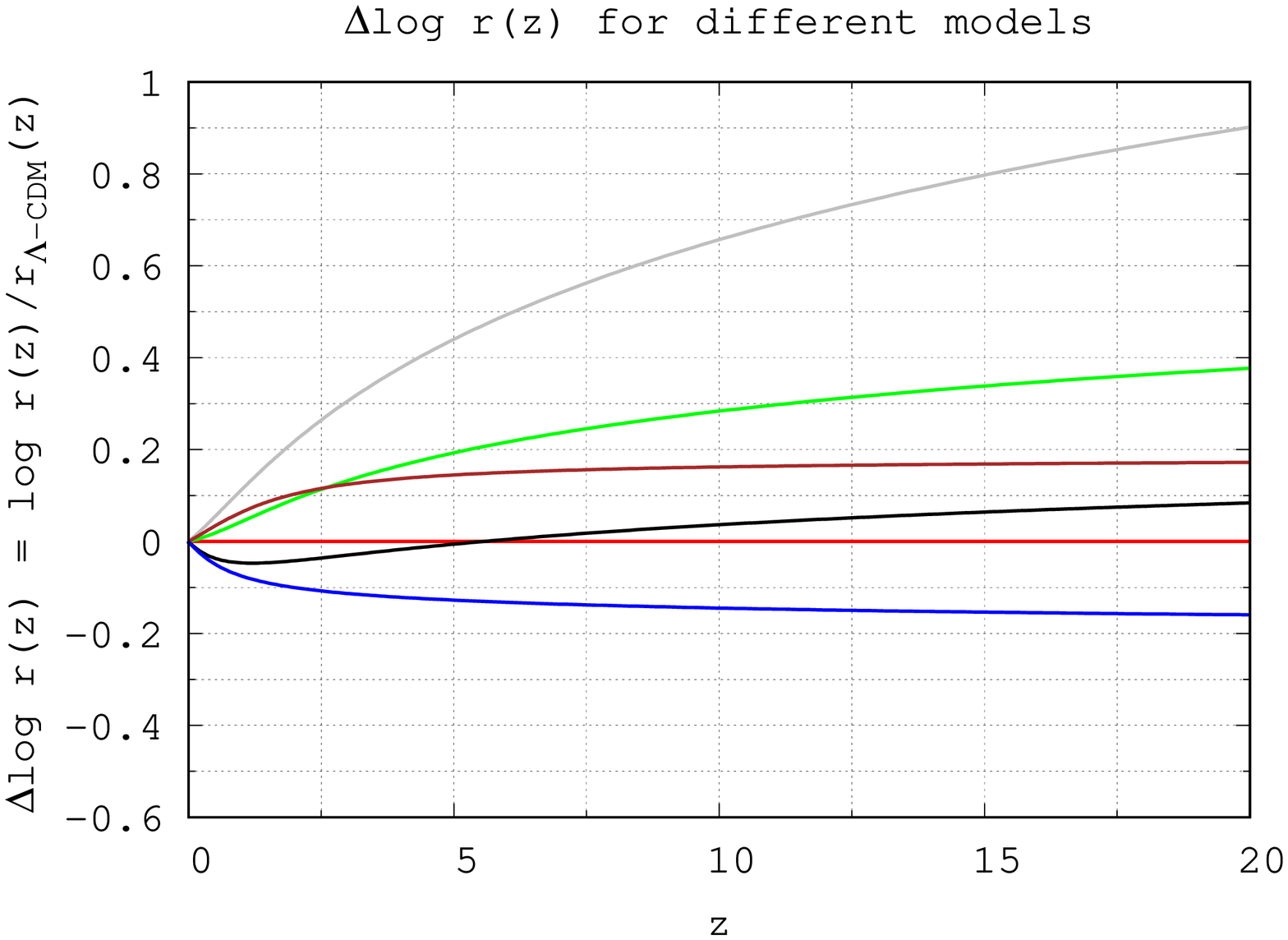}    
    \vfill
    \includegraphics[height=0.49\textwidth,angle=-90]{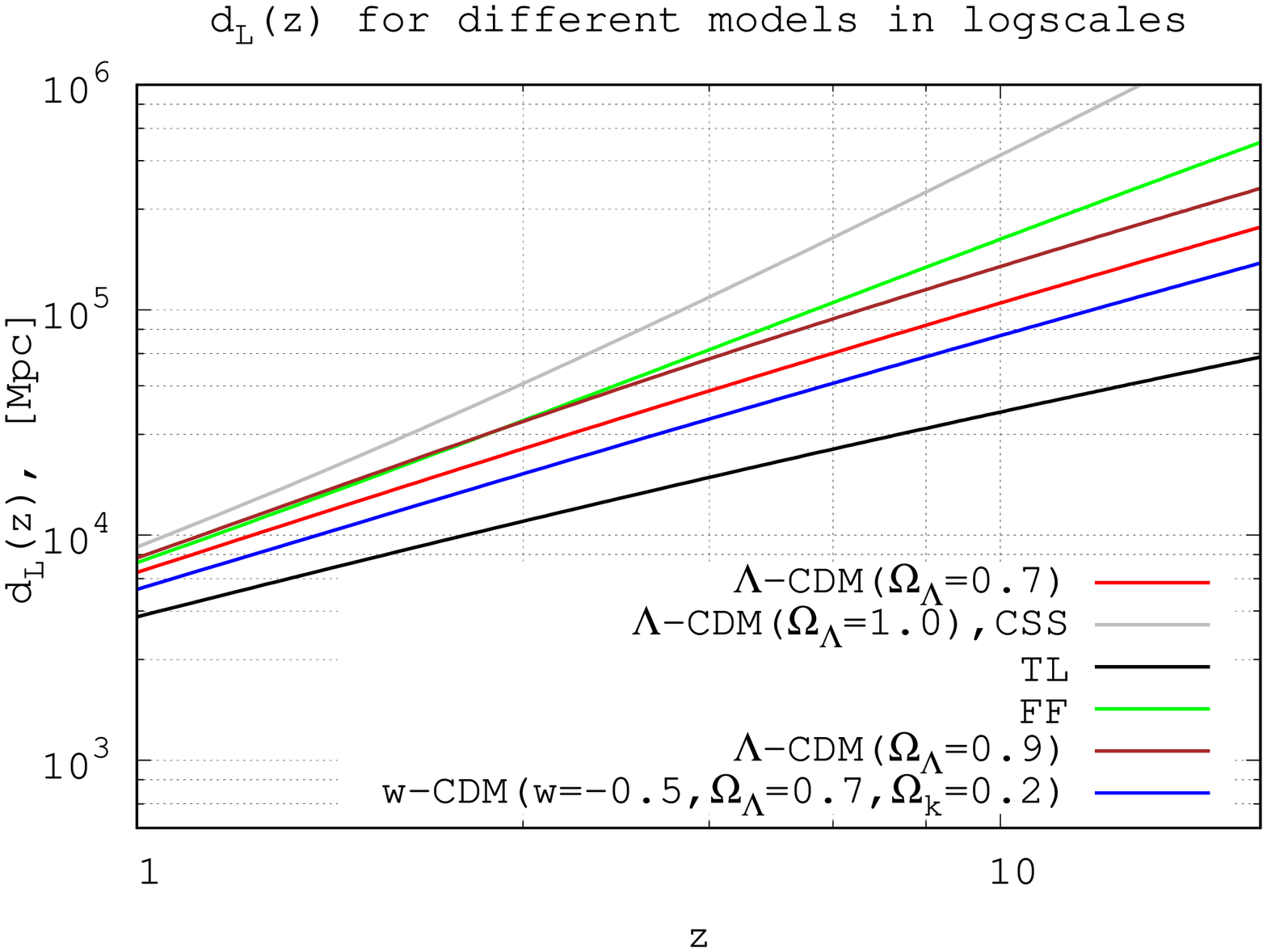}
    \hfill
    \includegraphics[height=0.49\textwidth,angle=-90]{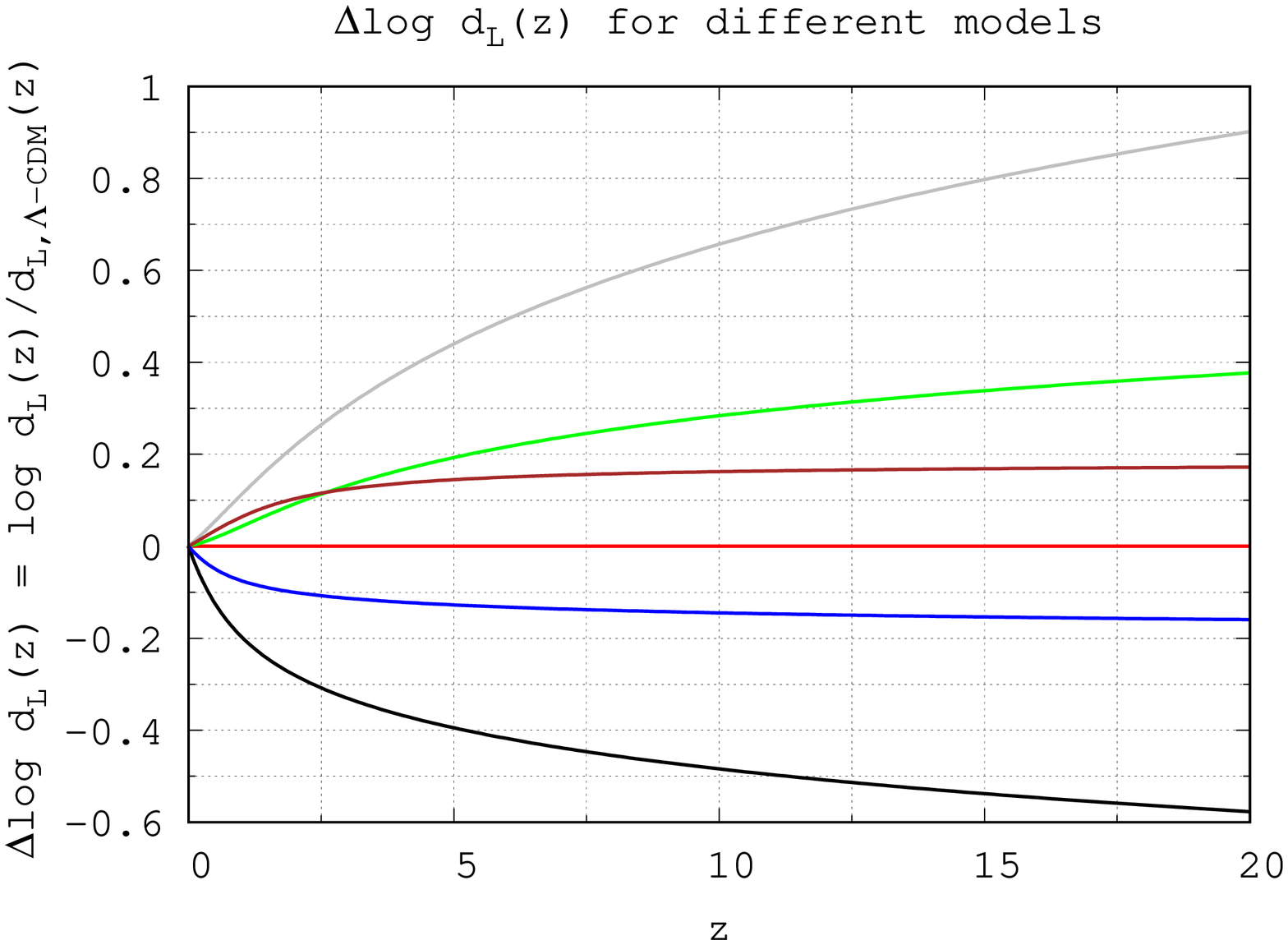}
    \vfill
    
    \caption{The metric distance  $r(z)$ (top)  and the luminosity distance $d_L(z)$ (bottom) for six considered cosmological models described in Secs.\ref{standard-model-1} and \ref{models-param}. Left: direct distances, Right: residuals from the standard $\Lambda$CDM model.
    }
    \label{fig:dL}
\end{figure*}

The area of the sphere is $4\pi(S(t_0)\mu)^2=4\pi l^2(z)$.
According to Eq.~(\ref{flux}), for the source isotropic luminosity $L_\text{iso}$ we measure the bolometric flux $F_L$
at luminosity distance $d_L$ as
\begin{equation}
\label{flux-lum-L}
  F_L\left[\frac{\text{erg}}{\text{s~cm}^2}\right]
  = \frac{L_\text{iso}} {4 \pi l^\text{2}(z) (1+z)^\text{2}}
  = \frac{L_\text{iso}}{4\pi d_L^2(z)}\,.    
\end{equation}
Here two factors $(1+z)$ take into account the observed energy decrease and time dilation effects. Hence the luminosity distance $d_L$ in the Friedmann's models is defined  according to Eq.~(\ref{l-z}) as
\begin{equation}
  d_L(z)=(1+z)\,l(z)=
  \frac{c}{H_0}
  \frac{(1+z)}{\sqrt{|\Omega^0_\text{k}|}}I_{k}
  \left(  \sqrt{|\Omega^0_\text{k}|}
  \int_0^z \frac{\text{d}z'}{h(z')} \right)\,,
    \label{lum-dist-L}
\end{equation}
where $\Omega^0_{k}$ is the curvature density parameter, and $I_{k}(x)=\sinh(x)$ for $\Omega_{k}<0, I_{k}(x)=x$ for $\Omega_{k}=0$, and $I_{k}(x)=\sin(x)$ for $\Omega_{k}>0$.  The metric and luminosity distances for the models from section 2.3.4 are given by equations (\ref{r-z}), (\ref{l-z}), and (\ref{lum-dist-L}) and represented in Fig.~\ref{fig:dL}.

For the total source bolometric energy $E_\text{iso}$ (erg) we measure the fluence $F_E$ (erg\,cm$^{-2})$ at ``energy distance'' $d_E$, so
we get from Eq.~(\ref{flux-E}),
\begin{equation}
\label{flux-lum-E}
  F_E\left[\frac{\text{erg}}{\text{cm}^2}\right]
  = \frac{E_\text{iso}} {4 \pi l^\text{2}(z) (1+z)}
  = \frac{E_\text{iso}}{4\pi d_E^2(z)}\,,    
\end{equation}
where the factor $(1+z)$ takes into account the energy decreases, and from  Eq.~(\ref{l-z}) we get the expression for the energy distance in the form
\begin{equation}
  d_E(z)=\sqrt{(1+z)}l(z)=
  \frac{c}{H_0}
  \frac{\sqrt{(1+z)}}
  {\sqrt{|\Omega^0_{k}|}}
  I_{k}
  \left(  \sqrt{|\Omega^0_{k}|}
  \int_0^z \frac{\text{d}z'}{h(z')} \right)\,.
    \label{lum-dist-E}
\end{equation}
The Hubble radius is
$R_{{H}_0}=c/H_0 \approx 4\,286\,h^{-1}_\text{70}$ Mpc.
According to recent Planck results, which probe the redshift $z \sim 1\,000$, the primary cosmological parameters are the curvature parameter $\Omega_{k} = -0.0007\pm0.0012$
(in our definition $\Omega_{k}$ has the same  sign as the curvature constant $k$), the EoS parameter 
$w= -1.03 \pm 0.03$, and the Hubble parameter
$H_0 = 67.4 \pm 0.5$ km~s$^{-1}$~Mpc$^{-1}$~\citep{Aghanim2018}. 

However, recent local measurements of the HD   give the  $H_0 =74.03 \pm 1.42$ km~s$^{-1}$ Mpc$^{-1}$, which points to some new physics beyond $\Lambda$CDM~\citep{Riess2019}. So it is important to consider HD for different cosmological models at intermediate  redshifts up to $z \sim 20$, achievable by LGRB.
In our calculations we shall use the standard value of the Hubble parameter 
$H_0 =70$ km~s$^{-1}$ Mpc$^{-1}$.

\section{Hubble Diagram in logarithmic scale for considered cosmological models }

\begin{figure*}
    \includegraphics
        [height=0.49\textwidth,angle=-90]   
        {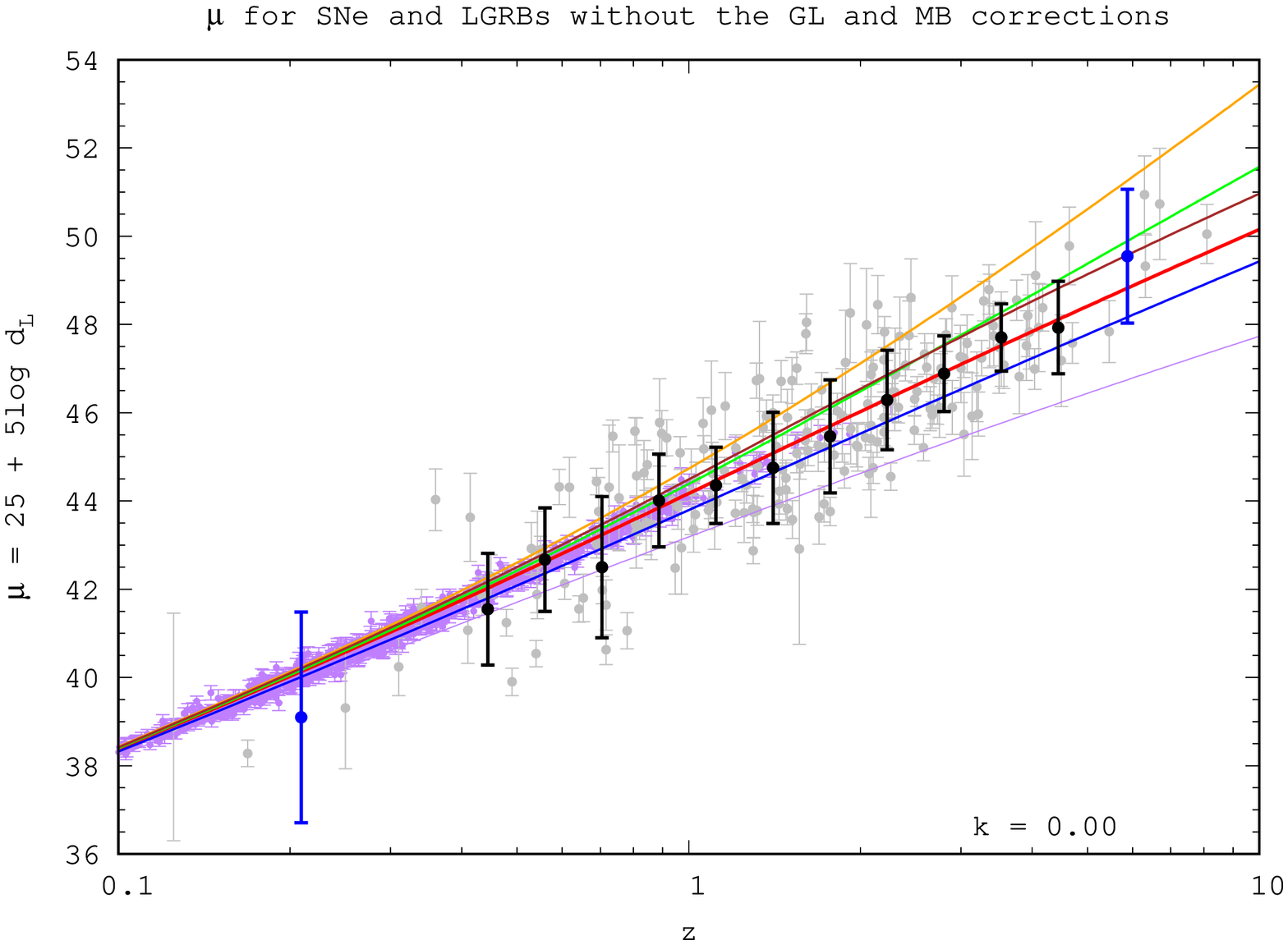}
    \hfill
    \includegraphics
        [height=0.49\textwidth,angle=-90]
        {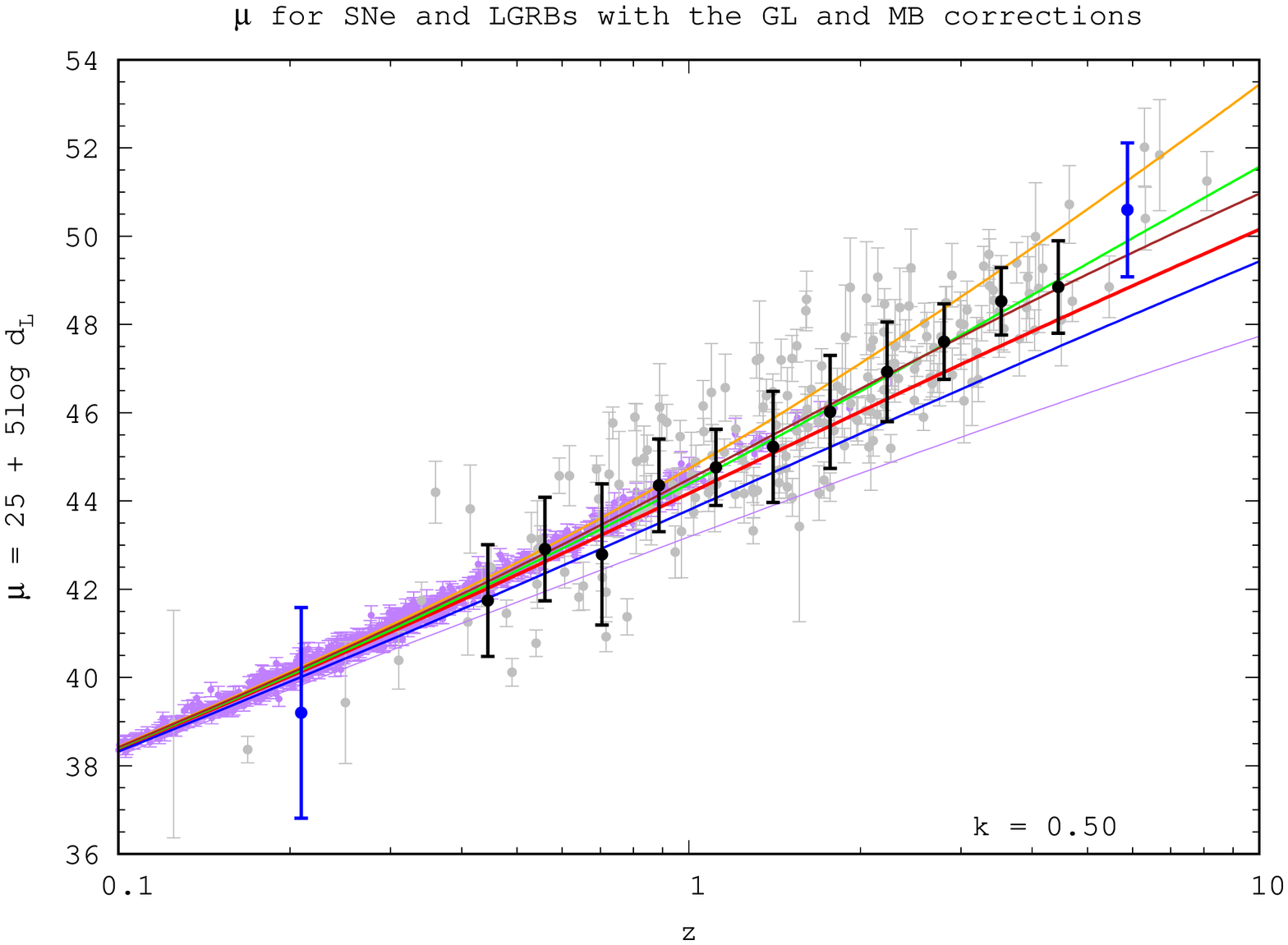}
    \vfill 
    \includegraphics    
        [height=0.49\textwidth,angle=-90]   
        {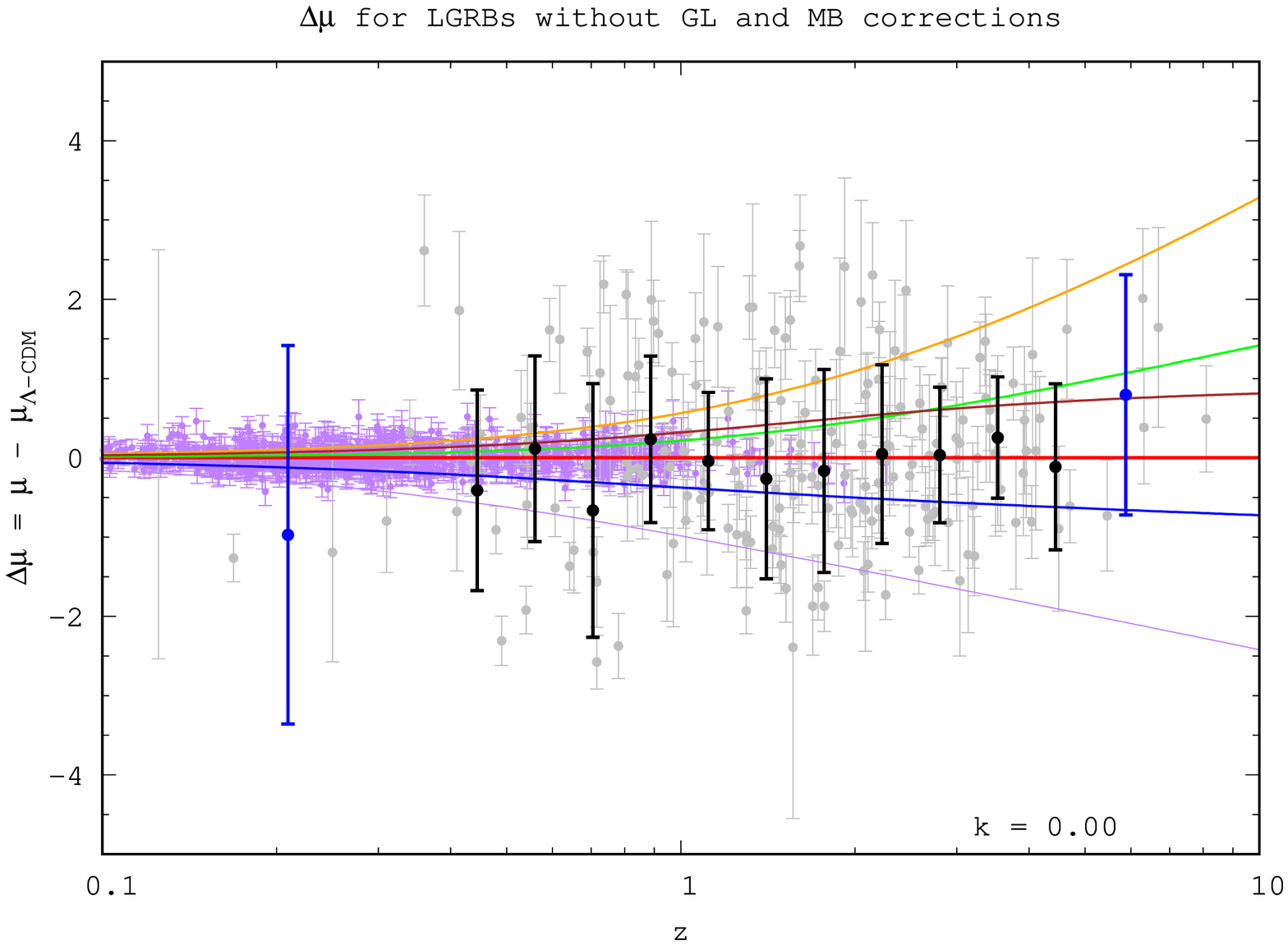}
    \hfill
    \includegraphics
        [height=0.49\textwidth,angle=-90]
        {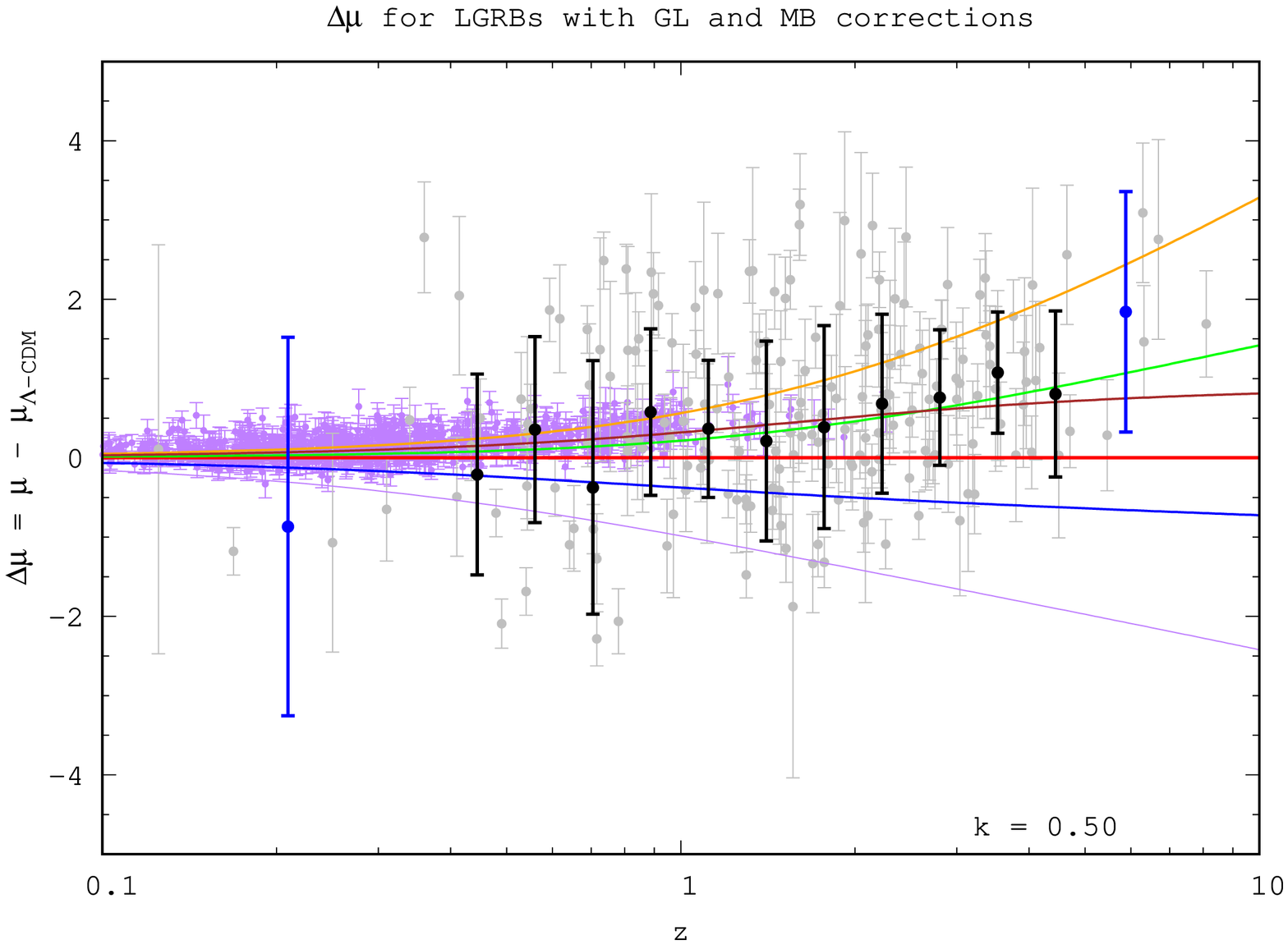}
    \caption{
        \emph{Top panels}: the luminosity distance modulus $\mu$ versus redshift  (HD) in logarithmic  $z$-scale for the SN Ia and LGRB samples. Black points are  the  median values of $\mu$ with logarithmic step  $\Delta \log{z}=0.1$ of the LGRB sample.
        \emph{Bottom panels}: the residuals $\Delta \mu$  from the standard $\Lambda$CDM model   for the observed luminosity distance modulus. \emph{Left}: without the correction for the GLB and MB. \emph{Right}: corrected with $k=0.5$.
        The colour curves correspond to the  cosmological models defined in Section~\ref{Hubble Diagram in Cosmological Models}.  
    }
    \label{fig7:HD193log}
\end{figure*}

Figure~\ref{fig7:HD193log} presents the HD  in logarithmic scale and with the logarithmic scale median values ($\Delta \log{z}=0.1$) of the LGRB sample. The blue points are  median values at $z<0.35$ and $4.2<z$, respectively. The points and curves are the same as in Fig.\ref{fig6:HD193lin}.

\bsp	
\label{lastpage}
\end{document}